\documentclass[12pt]{article}

\usepackage{a4wide}
\usepackage{cite}
\usepackage{amssymb}
\usepackage{amsmath}
\usepackage{epsfig}
\usepackage{eucal}

\newlength{\horizMargin}
\newcommand{\as}{\alpha_\mathrm{s}}      
\newcommand{\asb}{\bar{\alpha}_\mathrm{s}}             
\newcommand{\asbmu}{\asb}              
\newcommand{\ave}[1]{\left\langle #1 \right\rangle} 
\newcommand{\chie}{\chi_{\e}}            
\newcommand{\cotg}{\mathrm{cotg}}        
\newcommand{\DIS}{\mathrm{DIS}}          
\newcommand{\dif}{\mathrm{d}}            
\newcommand{\e}{\varepsilon}
\newcommand{\esp}[1]{\mathrm{e}^{#1}}    
\newcommand{\ff}{\mathcal{N}}              
\newcommand{\gGG}{\gamma_{\pg\pg}}
\newcommand{\gQG}{\gamma_{\pq\pg}}
\newcommand{\half}{{\textstyle\frac{1}{2}}}
\newcommand{\Kernel}{\mathcal{K}_{\e}}   
\newcommand{\kt}{\boldsymbol{k}}          
\newcommand{\lag}{{\CMcal L}_{\e}}       
\newcommand{\lt}{\boldsymbol{l}}          
\newcommand{\MSbar}{\overline{\mathrm{MS}}}

\newcommand{\om}{\omega}
\newcommand{\ord}[1]{\CMcal{O}\left(#1\right)}
\newcommand{\pg}{\mathsf{g}}             
\newcommand{\pq}{\mathsf{q}}             
\newcommand{\pt}{\boldsymbol{p}}          
\newcommand{\qKer}{H}                    
\newcommand{\qKernel}{\qKer_{\e}^{(p)}}  
\newcommand{\qRes}{\mathcal{H}}          
\newcommand{\qResR}{\qRes^{\mathrm{rat}}}
\newcommand{\qResT}{\qRes^{\mathrm{tran}}}
\newcommand{\qcf}[1][p]{\qRes_{\e}^{(#1)}} 
\newcommand{\rk}{\mathfrak{R}}           
\newcommand{\sign}{\mathrm{sign}}        
\newcommand{\ugd}{{\CMcal F}}            
\newcommand{\ui}{\mathrm{i}}             

\numberwithin{equation}{section}

\begin{document}


\titlepage

\begin{flushright}
  DFF 426/07/05\\
  hep-ph/0507106 \\
\end{flushright}

\vspace*{1in}

\begin{center}
  {\Large \bf
  Dimensional Regularisation and Factorisation Schemes\\[3mm]
  in the BFKL Equation at Subleading Level}
  
  \vspace*{0.4in}
  
  M.~Ciafaloni and D.~Colferai
  
  {\small
  \vspace*{0.5cm}
  {\it  Dipartimento di Fisica, Universit\`a di Firenze,
   50019 Sesto Fiorentino (FI), Italy}; \\
  \vskip 2mm
  {\it  INFN Sezione di Firenze,  50019 Sesto Fiorentino (FI), Italy}\\
  \vskip 2mm}
\end{center}


\bigskip

\begin{abstract}
  \noindent
  We study the anomalous dimensions and coefficient functions generated by the
  BFKL equation in $4+2\e$ dimensions, by investigating both running coupling
  effects, and the inclusion of the full next-to-leading kernel. After
  generalising the Fourier representation of the solutions to this case, we
  analyse the $\beta$-dependent renormalisation-group factorisation of anomalous
  dimension and coefficient contributions to the gluon density. We derive on
  this basis the normalisation factor of the $Q_0$-scheme with respect to the
  $\MSbar$-scheme, including $\beta$-dependent corrections to it, and we outline
  the derivation of the full next-to-leading contributions. We also provide an
  expression for the resummed $\gQG$ in the $\MSbar$-scheme which exhibits its
  universality and is explicit up to quadratures.
\end{abstract}

\newpage

\section{Introduction\label{s:intro}}

The relationship of the BFKL equation~\cite{BFKL} --- describing small-$x$
evolution in QCD --- with the DGLAP equation~\cite{DGLAP} --- describing $Q^2$
evolution --- has been the subject of several investigations in the
nineties~\cite{Kfact,CoEl91,CaCiHa93,CaHa94,RGvert,QQvertCC,QQvertFFFK}, in the
attempt of providing a unified picture of small-$x$ physics in QCD. It is well
known that consistency of the BFKL approach with renormalisation group (RG)
factorisation is achieved by means of resummation formulas of large
contributions at small-$x$ to the quark and gluon anomalous dimensions and
coefficient functions, which have been derived at the leading-$\log x$ (LL$x$)
level~\cite{Kfact} and, in part, at the next-to-leading (NL$x$)
one~\cite{CaCiHa93,CaHa94}.

In particular, Catani and Hautmann~\cite{CaHa94} have studied the LL$x$ BFKL
equation in $4+2\e$ dimensions with dimensional regularisation (and frozen
coupling) and have derived, on this basis, the resummation of leading
logarithms~\cite{CaCiHa93,CaHa94} for the gluon coefficient function --- usually
called $R(\as)$ --- and for the quark anomalous dimension $\gQG$ in the
$\MSbar$- and DIS-schemes. This calculation extends in part into the NL$x$
level, not only because of $\gQG$ being next-to-leading, but also because of the
$R$-factor, which provides NL$x$ corrections to the gluon
anomalous dimension $\gGG$ when running coupling effects are turned on~\cite{CaCi97}.

It is known, on the other hand, that full NL$x$ terms~\cite{FaLi98,CaCi98} are
quite large and negative, and that doubly resummed
approaches~\cite{CCS,CCSSkernel,ABF} are required in order to stabilise the
subleading-log series. This raises the question of a better analysis of
anomalous dimensions and coefficient functions at subleading level in various
factorisation schemes. While the $\kt$-factorisation schemes (like the
$Q_0$-scheme~\cite{Q0}, characterised by an off-shell initial parton) have been
pushed to the doubly-resummed level~\cite{CCSSkernel}, the minimal subtraction one
(characterised by dimensional regularisation, and normally used in fixed order
calculations) needs yet to be extended to a full treatment of NL$x$ terms.

The purpose of the present paper is to devise a method to perform such analysis,
and to work it out in detail in the case of the BFKL equation with running
coupling, which already contains quite important subleading effects. The method
is then extended to a full treatment of NL$x$ coefficient terms, which require
the $\e$-dependence of the NL$x$ BFKL kernel at least up to $\ord{\e}$.
Unfortunately, the latter is yet to be extracted from the
literature~\cite{RGvert,QQvertCC,QQvertFFFK}.

The main tool of our analysis is the generalisation to $4+2\e$ dimensions of the
$\gamma$-representation of the gluon density --- that is, of a Fourier
representation of the BFKL solution in which $\gamma$ is conjugate to
$t\equiv\log(\kt^2/\mu^2)$. While for $\e=0$ the running-coupling equation is a
differential equation in $\gamma$, for $\e\neq0$ it becomes a finite-difference
equation which is treated in Secs.~\ref{s:grLL} and \ref{s:rce} and in
App.~\ref{a:sde}. This allows one to write the gluon density in a generalised
$Q_0$-scheme%
\footnote{The label $Q_0$ referred originally~\cite{Q0} to the fact that the
  initial gluon, defined by $\kt$-factorisation, was set off-mass-shell ($\kt^2
  = Q_0^2$) in order to cutoff the infrared singularities. It will be shown,
  however, that the effective anomalous dimension at scale $\kt^2 \gg Q_0^2$ is
  independent of the cut-off procedure, whether of dimensional type or of
  off-mass-shell one.}
as the product of an anomalous dimension exponential and of a fluctuation factor
that we call $\ff$. We then show in Sec.~\ref{s:grLL} that the factor
$\rk\equiv R/\ff$ is due to the $\ord{\e}$ dependence of the LL$x$ gluon
anomalous dimension. This result offers an interpretation of the mismatch of $R$
and $\ff$ coefficients, and a hint to possible generalisations, investigated in
Secs.~\ref{s:rce}-\ref{s:nlc}.

The general case with running coupling ($b>0$) is treated in Sec.~\ref{s:rce},
and is qualitatively similar to the frozen coupling ($b=0$) case, except that
the beta-function
\begin{equation}\label{beta}
 \beta(\asb) \equiv \frac{\dif\asb(t)}{\dif t} = \e \asb - b \asb^2 + \ord{\asb^3}
\end{equation}
has both the dimensional contribution ($\e$-term) and the running-coupling one
($b$-term). We are thus able to explain how RG factorisation works for
coefficient and anomalous dimension parts for $b>0$, thus relating the
$Q_0$-scheme (with dimensional regularisation) and the $\MSbar$-scheme in an
unambiguous way. In particular, we confirm that the two anomalous dimensions
differ by the quantity $\dif\log R / \dif t$.

Secs.~\ref{s:fsl} and \ref{s:nlc} are devoted to the calculation of the NL$x$
corrections to $R$. In Sec.~\ref{s:fsl} we treat the corrections due to the
running coupling, which require the (known) $\e$-dependence of the leading
kernel eigenvalue up to $\ord{\e^2}$, and the corresponding refinement of the
$\gamma$-representation and of the saddle-point fluctuation formalism. In
Sec.~\ref{s:nlc} we calculate the remaining NL$x$ corrections, due to the
inclusion of the NL$x$ kernel. Here the finite difference equation involves two
steps, and is truncated at NL$x$ level by an iterative procedure. The final
result involves the $\ord{\e}$ corrections to the NL$x$ kernel eigenvalue,
which are not yet explicitly known.

In Sec.~\ref{s:ugqg} we reconsider the calculation of $\gQG$ in the
$\MSbar$-scheme.  Contrary to the case of the DIS-scheme, in which the
coefficient function $C_{\pq\pg}$ is set to zero, no closed form resummation
formula is yet available for $\gQG^{(\MSbar)}$. Catani and Hautmann are able to
provide a recursive method in order to disentangle $C_{\pq\pg}$ from $\gQG$, so
as to calculate a number of terms of their expansion in $\asb/\om$, $\om$ being
the Mellin moment conjugate to $x$, a result further improved in the later
literature~\cite{Rterms}. Here we use the $\gamma$-representation of the gluon
density in order to do the same calculation: this allows us to provide an
explicit resummation formula which exhibits the universality of $\gQG$ and
involves only quadratures of functions which are known in principle. However,
the latter calculation requires an all-order $\e$-expansion of the coefficient
function of the gluon. An interesting byproduct of the universality analysis is
the proof (App.~\ref{a:cfqk}) that the {\em off-shell} $P_{\pq\pg}$ splitting
function introduced in~\cite{CaHa94} is indeed process-independent.

In the final Sec.~\ref{s:disc} we summarise our results and we discuss their
relevance for the improved resummations of splitting
functions~\cite{CCSSkernel,ABF}.

\section{$\boldsymbol\gamma$-representation and $\boldsymbol R$-factor
in the LL$\boldsymbol x$ case\label{s:grLL}}

Investigating the BFKL equation in $4+2\e$ dimensions is of importance on its
own, because --- in addition to the ultraviolet (UV) role of dimensional
regularisation --- a positive $\e$ parameter allows to regularise the mass-shell
singularities and, in some regime (see Sec.~\ref{s:rce}) it avoids the Landau
pole. For this reason we shall consider $\e>0$ as an infrared cutoff, in
alternative to setting off-mass-shell the initial condition for partonic
evolution, as done in the $Q_0$-scheme~\cite{Q0}. Our purpose is, at
large, to understand the relationship of such two kinds of schemes, whether or
not a minimal subtraction is assumed in the partonic densities.

Since the gauge coupling $g$ is dimensionful in $4+2\e$ dimensions, we shall
introduce, as usual, the renormalisation scale $\mu$, the dimensionless coupling
in the $\MSbar$-scheme
\begin{equation}\label{d:alphas}
 \as \equiv \frac{(g\mu^{\e})^2}{(4\pi)^{1+\e} \esp{\e\psi(1)}} \;,
 \quad \asb \equiv \as \frac{N_c}{\pi} \;,
\end{equation}
and the parameter $t\equiv\log(\kt^2/\mu^2)$ in terms of which the running
coupling is $\asb(t) \equiv \asb \esp{\e t}$. Therefore, the LL$x$ equation
shows a running coupling with infrared free evolution corresponding to the
dimensional contribution of the beta function (\ref{beta}) in the $b\to0$ limit.
Since $\asb(t)\to0$ for $t\to-\infty$, we can set on-shell ($\kt_0=0$) the
initial condition for the gluon Green's function, so that $\e$ acquires the role
of infrared cutoff.  The ensuing solution of the LL$x$ BFKL equation is
naturally expressed as a power series in $\asb(t)$, which was extensively studied
in~\cite{CaHa94}.

Our purpose here is to recast the solution for the gluon density mentioned above
in the $\gamma$-representation form --- $\gamma$ being conjugate to $t$ --- so
as to be able to describe in a simpler way its anomalous dimension behaviour.
The BFKL equation for the unintegrated gluon density $\ugd$ in $4+2\e$
dimensions reads (the $\om$-dependence of $\ugd$ is understood)
\begin{align}
 \ugd_{\e}(\kt) &= \delta^{(2+2\e)}(\kt) + \frac1{\om}
 \int\frac{\dif^{2+2\e}\kt'}{(2\pi)^{2+2\e}}\;\Kernel(\kt,\kt')\ugd_{\e}(\kt')
 \label{BFKLeq} \\
 &= \delta^{(2+2\e)}(\kt) + \frac{\esp{\e\psi(1)}}{(\pi\kt^2)^{1+\e}}
 \widetilde{\ugd}_{\e}(\kt) \;,
 \nonumber \\
 \Kernel(\kt,\kt') &= g^2 N_c \left[\frac1{\pi(\kt-\kt')^2} - (\pi\kt^2)^{\e}
 \frac{\Gamma^2(1+\e)\Gamma(1-\e)}{\e\Gamma(1+2\e)} \delta^{2+2\e}(\kt-\kt')
 \right]
 \label{Kernel}
\end{align}
and its power series solution is
\begin{equation}\label{powerSol}
 \widetilde{\ugd}_{\e}(\kt) = \frac{\asb(t)}{\om}
 \left[1+\sum_{n=1}^\infty \left(\frac{\asb(t)}{\om}\right)^n
 \prod_{k=1}^n \chi_{\e}(k\e)\right] \;,
\end{equation}
where
\begin{equation}\label{chie}
 \chie(\gamma) = \frac{\esp{\e\psi(1)}\Gamma(1+\e)}{\e} \left[
 \frac{\Gamma(\gamma)\Gamma(1-\gamma)}{\Gamma(\gamma+\e)\Gamma(1-\gamma+\e)}
 - \frac{\Gamma(1+\e)\Gamma(1-\e)}{\Gamma(1+2\e)} \right]
\end{equation}
is the LL$x$ ``characteristic function'' of $\Kernel$ in $4+2\e$ dimensions,
defined by
\begin{equation}\label{d:chie}
  \int\frac{\dif^{2+2\e}\kt'}{(2\pi)^{2+2\e}}\; \Kernel(\kt,\kt')
  (\kt'{}^2)^{\gamma-1-\e} \equiv \asbmu \chie(\gamma)
  \frac{(\kt^2)^{\gamma-1}}{\mu^{2\e}} \;.
\end{equation}

The series (\ref{powerSol}) is well behaved in the infrared ($t\to-\infty$)
apart from the $1/\e$ poles, but is not really suitable in the ultraviolet
($t\to+\infty$), where the variable $\asb(t)/\om$ grows, and at most a finite
radius of convergence is expected. We thus look for a solution in the
$\gamma$-representation form
\begin{equation}\label{d:gammaRep}
 \widetilde{\ugd}_{\e}(\kt) = \int_{\Re\gamma=c}\frac{\dif\gamma}{2\pi\ui} \;
 \esp{\gamma t} f_{\e}(\gamma) \;,
\end{equation}
where the Fourier variable $\gamma$ has the interpretation of a continuum in
which $\gamma=n\e$ is the lattice counterpart. We expect Eq.~(\ref{d:gammaRep}) to
be best suited for anomalous dimension properties in which $n\to\infty$ and
$\e\to0$ with $\gamma=n\e$ kept fixed.

Let us start noticing that the ``ansatz''~(\ref{d:gammaRep}), when replaced in
the BFKL equation with a general initial condition
$\displaystyle{f_{\e}^{(0)}(\gamma)}$, leads to a finite difference equation of
the form
\begin{equation}\label{fde}
 f_{\e}(\gamma+\e) = f_{\e}^{(0)}(\gamma+\e)
 + \frac{\asbmu}{\om} \chie(\gamma) f_{\e}(\gamma) \;.
\end{equation}
In the following we shall often consider the case
\begin{equation}\label{f0}
 f_{\e}^{(0)}(\gamma+\e) = f^{(0)}(\gamma) \equiv \frac{\asbmu}{\om}
 \frac{\esp{\gamma T}}{\gamma} \;,
\end{equation}
which corresponds to the initial condition
\begin{equation}\label{F0}
 \widetilde{\ugd}_{\e}^{(0)}(\kt) = \frac{\asbmu \esp{\e t}}{\om} \Theta(t+T) \;,
\end{equation}
i.e.\ to the one expected from Eq.~(\ref{powerSol}) with a
cutoff at $\kt^2 = Q_0^2 \equiv \mu^2 \esp{-T}$. In this way we can
study the role of the cutoff, by solving (\ref{fde}) with $T$ fixed, and
performing the limits $T\to +\infty$, $\e\to0$ either in this order, or in
reverse order.

We shall solve Eq.~(\ref{fde}) in two steps, starting from the homogeneous
equation
\begin{equation}\label{homEq}
 h_{\e}(\gamma+\e) = \frac{\asbmu}{\om} \chie(\gamma) h_{\e}(\gamma)
 \equiv \esp{L_{\e}(\gamma)} h_{\e}(\gamma) \;,
\end{equation}
where we take the ansatz
\begin{equation}\label{hForm}
 h_{\e}(\gamma) = \exp\left\{\int^{\gamma} \lag(\gamma')\frac{\dif\gamma'}{\e}
 \right\} \equiv \exp\big\{S_{\e}(\gamma)\big\}
\end{equation}
so that Eq.~(\ref{homEq}) implies
\begin{equation}\label{lagEq}
 S_{\e}(\gamma+\e) - S_{\e}(\gamma) =
 \int_{\gamma}^{\gamma+\e} \lag(\gamma')\frac{\dif\gamma'}{\e}
 = L_{\e}(\gamma) \;.
\end{equation}
We show in App.~\ref{aa:as} that the solution for the ``Lagrangian''
$\lag(\gamma)$ is expressed, under appropriate conditions, in terms of the
Bernoulli numbers $B_n$ as follows
\begin{equation}\label{lagExp}
 \lag(\gamma) = \sum_{n=0}^{\infty} \frac{B_n}{n!} \e^n L_{\e}^{(n)}(\gamma) \;,
\end{equation}
where $L^{(n)}$ is the $n$-th derivative of $L$ with respect to $\gamma$, and the
generating function is
\begin{equation}\label{bernoulli}
 \sum_{n=0}^{\infty} \frac{B_n}{n!} z^n = \frac{z}{\esp{z}-1} \;,
\end{equation}
so that
\begin{equation}\label{Bn}
 B_0=1, \quad B_1=-\frac1{2}, \quad B_2 = \frac1{6},
 \quad B_3 = 0, \quad B_4 = -\frac1{30}, \cdots \;.
\end{equation}
Correspondingly the ``action'' takes the form
\begin{equation}\label{action}
 S_{\e}(\gamma) = \frac1{\e}\int^{\gamma} L_{\e}(\gamma') \dif\gamma'
 -\frac1{2} L_{\e}(\gamma) + \frac{\e}{12} L_{\e}'(\gamma) + \ord{\e^2} \;,
\end{equation}
where $L_{\e}$ can be further expanded in $\e$ as follows
\begin{equation}\label{Lexp}
 L_{\e}(\gamma) \equiv \log\Big(\frac{\asbmu}{\om}\chie(\gamma)\Big) =
 \log\Big(\frac{\asbmu}{\om}\chi_0(\gamma)\Big)
 + \e \frac{\chi_1(\gamma)}{\chi_0(\gamma)}
 + \e^2 \left( \frac{\chi_2(\gamma)}{\chi_0(\gamma)}
   - \frac12 \frac{\chi_1^2(\gamma)}{\chi_0^2(\gamma)} \right)
 + \ord{\e^3}
\end{equation}
and the BFKL eigenvalue function $\chie = \chi_0 + \e\chi_1 + \e^2\chi_2 + \ord{\e^3}$ is
given, according to Eq.~(\ref{chie}), by
\begin{align}
 \chi_0(\gamma) &= 2\psi(1)-\psi(\gamma)-\psi(1-\gamma)
 \label{chi0} \\
 \chi_1(\gamma)
 &= \frac12 \left[ \chi_0^2(\gamma) +
 2\psi'(1)-\psi'(\gamma)-\psi'(1-\gamma) \right]
 \label{chi1} \\
 \chi_2(\gamma)
 &= \chi_0(\gamma) \big[ \chi_1(\gamma) -\frac12\psi'(1)\big]
 -\frac13 \chi_0^3(\gamma) + \frac16 \big[ 8\psi''(1)-\psi''(\gamma)
 -\psi''(1-\gamma) \big] \;.
 \label{chi2}
\end{align}
Therefore, the solution of the homogeneous equation in $t$-space
(Eq.~(\ref{BFKLeq}) without delta term) takes the form
\begin{equation}\label{solHomEq}
 \widetilde{\ugd}_{\e}(\kt) = \int\dif \gamma\;
 \frac{\esp{\gamma t}}{\sqrt{\chi_0(\gamma)}}
 \exp\left\{\frac1{\e}\int^{\gamma}
 \log\Big(\frac{\asbmu}{\om}\chi_0(\gamma')\Big) \dif\gamma'
 +\int^{\gamma} \frac{\chi_1(\gamma')}{\chi_0(\gamma')} \dif\gamma' \right\}
 \times \big[1+\ord{\e}\big] \;,
\end{equation}
where we have truncated the $\e$-expansion of the exponent up to the finite
terms. The choice of the lower bound of the $\gamma'$ integrals is of no
concern, since the normalisation of the solution of the homogeneous equation is
arbitrary.

The inhomogeneous equation (\ref{fde}) has, on the other hand, the iterative
solution
\begin{align}
 f_{\e}(\gamma) &= f^{(0)}(\gamma-\e)
 + f^{(0)}(\gamma-2\e) \frac{\asbmu}{\om}\chie(\gamma-\e) +
 f^{(0)}(\gamma-3\e)
 \frac{\asbmu}{\om}\chie(\gamma-2\e) \frac{\asbmu}{\om}\chie(\gamma-\e)
 + \cdots
\nonumber \\
 &= \sum_{n=1}^{\infty} f^{(0)}(\gamma-n\e)\left(\frac{\asbmu}{\om}
 \right)^{n-1}\prod_{m=1}^{n-1}\chie(\gamma-m\e)
 = \sum_{n=1}^{\infty} f^{(0)}(\gamma-n\e)
 \frac{h_{\e}(\gamma)}{h_{\e}(\gamma+\e-n\e)} \;,
\label{itSol}
\end{align}
which is closely related to the power series solution (\ref{powerSol}), as
analysed in more detail in App.~\ref{aa:is}. Here we just note the $\e=0$ limit of
Eq.~(\ref{itSol}) at fixed $T$. Due to the fact that $h_{\e}(\gamma)$ satisfies
Eq.~(\ref{homEq}), the ratios of $h$'s has a non trivial $\e\to0$ limit, and we
obtain
\begin{equation}\label{eps0lim}
 f_0(\gamma) = f^{(0)}(\gamma) \sum_{n=0}^{\infty}\esp{n L_0(\gamma)}
 = \frac{f^{(0)}(\gamma)}{1-\frac{\asbmu}{\om}\chi_0(\gamma)} \;,
\end{equation}
as expected for the solution of the BFKL equation with frozen coupling and with
cutoff. We get in fact, up to higher twist corrections,
\begin{equation}\label{frozenSol}
 \widetilde{\ugd}_0(\kt)
 = \frac{\esp{\gamma_0 T}}{-\gamma_0\,\chi_0'(\gamma_0)}
 \left(\frac{\kt^2}{\mu^2}\right)^{\gamma_0} \Theta(\kt^2-\mu^2\esp{-T}) \;,
\end{equation}
where $\gamma_0(\asbmu/\om)$ is the LL$x$ anomalous dimension, defined by
\begin{equation}\label{d:gamma0}
 1=\frac{\asbmu}{\om}\chi_0(\gamma_0) \;, \qquad
 \gamma_0\Big(\frac{\asbmu}{\om}\Big) = \frac{\asbmu}{\om} +
 \ord{\frac{\asbmu}{\om}}^4 \;.
\end{equation}
The result (\ref{frozenSol}) shows the customary infrared singular
$T$-dependence of the $Q_0$-scheme, determined by $\gamma_0$.

Our emphasis in this paper is, however, in the limit of (\ref{itSol}) for
$\e \ll 1$ and $T\to+\infty$, where we obtain its continuum counterpart which
roughly replaces the sum in the r.h.s.\ of (\ref{itSol}) with an integral, apart
from some $\e$-dependent corrections. In order to specify such corrections, we
rewrite Eq.~(\ref{fde}) in terms of
$\rho_{\e}(\gamma)\equiv f_{\e}(\gamma)/h_{\e}(\gamma)$ in the simpler form
\begin{equation}\label{rhoEq}
 \rho(\gamma+\e)-\rho(\gamma) = \frac{f^{(0)}(\gamma)}{h_{\e}(\gamma+\e)}
 = \frac{f^{(0)}(\gamma)}{h_{\e}(\gamma)} \esp{-L_{\e}(\gamma)} \;,
\end{equation}
which is of the type (\ref{lagEq}) and is solved in terms of Bernoulli numbers
as in (\ref{lagExp}). The r.h.s.\ of (\ref{rhoEq}) has, however, $1/\e$
singularities in the exponent (cf.\ Eqs.~(\ref{hForm}) and (\ref{action})), and
its derivatives generate eventually a non trivial correction factor (see
App.~\ref{aa:rho}) having a finite $\e\to0$ limit, as follows:
\begin{equation}\label{rho}
 \rho_{\e}(\gamma) = \int^\gamma \frac{\dif\gamma'}{\e}
 \frac{f^{(0)}(\gamma')}{h_{\e}(\gamma'+\e)}
 \frac{L_0(\gamma')-\e T}{1-\esp{-L_0(\gamma')+\e T}}
 \times\big[1+\ord{\e}\big] \;,
\end{equation}
where we have kept terms $\sim\e T$, that we consider a number of order unity.

Finally, by replacing Eq.~(\ref{rho}) into (\ref{d:gammaRep}) we get the
solution
\begin{align}
 \widetilde{\ugd}_{\e}(\kt) = \int\frac{\dif\gamma}{2\pi\ui}
 \int^{\gamma'}\dif\gamma'\;
 &\frac{
       \exp\left\{
       \gamma t
       + \frac1{\e} \int_0^{\gamma}L_0(z)\dif z
       + \gamma' T
       - \frac1{\e} \int_0^{\gamma'}L_0(z)\dif z
       +\int_{\gamma'}^{\gamma}\frac{\chi_1(z)}{\chi_0(z)}\dif z \right\}
      }{
       \e \gamma' \sqrt{\chi_0(\gamma)} \sqrt{\chi_0(\gamma')}
      }
\nonumber \\
 &\times \frac{L_0(\gamma')-\e T}{1-\esp{-L_0(\gamma')+\e T}} \;,
\label{doubleGammaRep}
\end{align}
where we have truncated the $\e$-expansion to the finite terms. This solution
has the form of a double $\gamma$-representation, similarly to the customary
case $b\neq0$, $\e=0$, in which the $\e$-parameter replaces $b$ by providing an
infrared free running coupling.

Our goal is now to investigate the $T\to +\infty$ limit of
Eq.~(\ref{doubleGammaRep}) by removing the artificial infrared cutoff
$Q_0^2 = \mu^2 \esp{-T}$. In this limit the leading $\gamma'$-dependent phase is
generally large and given by
\begin{equation}\label{phase1}
 E(\gamma') = \gamma' T - \frac1{\e}\int_0^{\gamma'} L_0(z) \dif z \;,
\end{equation}
so that the $\gamma'$-integral is dominated by the saddle point $\bar{\gamma}'$:
\begin{equation}\label{saddle1}
 E'(\bar{\gamma}') = T - \frac1{\e} L_0(\bar{\gamma}') = 0 \;,
\end{equation}
which implies
\begin{equation}\label{gammabar1}
 1 = \frac{\asb(-T)}{\om}\chi_0(\bar{\gamma}') \;.
\end{equation}
Therefore, $\bar{\gamma}'\sim\asb(-T)/\om\to0$ for $T\to +\infty$
\footnote{The solution with $\bar{\gamma}' \simeq 1$ is relevant in the infrared
  region $\kt^2 \ll Q_0^2$ ($t \ll -T$), where it behaves as
  $\sim \kt^2/Q_0^2 = \esp{T+t}$.}
and, by taking into account also the $\gamma'$-fluctuations
\begin{equation}\label{fluct1}
 \sigma_{\gamma'} \equiv \frac1{\sqrt{E''(\bar{\gamma}')}}
 = \sqrt{\frac{\e\chi_0(\bar{\gamma}')}{-\chi'_0(\bar{\gamma}')}} \;,
\end{equation}
all $T$-dependent factors in (\ref{doubleGammaRep}) cancel out, yielding
\begin{equation}\label{gammaRep}
 \widetilde{\ugd}_{\e}(\kt) \stackrel{T\to +\infty}{\longrightarrow}
 \int\frac{\dif\gamma}{\sqrt{2\pi\e}} \frac1{\sqrt{\chi_0(\gamma)}}
 \exp\left\{ \gamma t+\frac1{\e}\int_0^{\gamma} L_0(\gamma')\dif\gamma'
 + \int_0^{\gamma} \frac{\chi_1(\gamma')}{\chi_0(\gamma')}\dif\gamma' + \ord{\e} \right\} \;,
\end{equation}
which is just a solution (\ref{solHomEq}) of the {\em homogeneous} equation with
an appropriate normalisation.

Eq.~(\ref{gammaRep}) is the main result of this section and shows the mechanism
by which the solution becomes independent of the details of the initial condition
in the $T\to +\infty$ ($Q_0\to0$) limit. In fact, due to the infinite
evolution path from $-\infty$ to $t$, the shape of the solution reduces to the
one of the homogeneous equation, and the initial condition determines only the
normalisation. This already suggests the factorisation of the $1/\e$
singularities which have replaced the $T$-dependence in Eq.~(\ref{gammaRep}). In
order to prove such factorisation we resort again to the saddle point method in
order to evaluate the $\e\to0$ behaviour of (\ref{gammaRep}). The stationarity
condition is now
\begin{equation}\label{saddle2}
 \e t + L_0(\bar{\gamma}) = 0 \quad \iff \quad 1
 = \frac{\asb(t)}{\om} \chi_0(\bar{\gamma}) \;,
\end{equation}
with a stable fluctuation along the real axis for $\chi_0'(\bar{\gamma}) < 0$,
thus yielding the result
($\bar{\gamma} \equiv \bar{\gamma}_t = \gamma_0\big(\asb(t)/\om\big)$)
\begin{equation}\label{ugdTilde}
 \widetilde{\ugd}_{\e}(\kt) = \frac1{\sqrt{-\chi_0'(\bar{\gamma}_t)}}
 \exp\left\{\bar{\gamma}_t\,t + \frac1{\e}\int_0^{\bar{\gamma}_t}
 L_0(\gamma')\dif\gamma' + \int_0^{\bar{\gamma}_t}
 \frac{\chi_1(\gamma')}{\chi_0(\gamma')}\dif\gamma'
 \right\} \times \big[1+\ord{\e}\big] \;.
\end{equation}
Finally, we calculate the integrated gluon density in the form
\begin{align}
 g_{\e}(t) &\equiv \int\dif^{2+2\e}\kt'\;\ugd_{\e}(\kt')\,\Theta(\kt^2-\kt'{}^2)
\nonumber \\
 &= \frac1{\bar{\gamma}_t\sqrt{-\chi_0'(\bar{\gamma}_t)}}
 \exp\left\{\bar{\gamma}_t \,t + \frac1{\e}\int_0^{\bar{\gamma}_t}
 L_0(\gamma')\dif\gamma' + \int_0^{\bar{\gamma}_t}
 \frac{\chi_1(\gamma')}{\chi_0(\gamma')}\dif\gamma'
 \right\} \times \big[1+\ord{\e}\big] \;,
\label{d:gluon}
\end{align}
where we have performed an integration by parts of $\widetilde{\ugd}$ in
(\ref{ugdTilde}).

Some remarks are in order. Firstly, the saddle point condition (\ref{saddle2})
provides a real positive anomalous dimension only in the perturbative regime in
which $\asb(t) < \om / \chi_0(\half)$, value at which the LL$x$ anomalous
dimension has the well known singularity at $\gamma=\half$. Since, however,
$\asb(t)$ increases with $t$ for $\e>0$, for large $t$ the solutions to
(\ref{saddle2}) will become necessary complex conjugate with
$\Re\bar{\gamma}=\half$, and $g_{\e}(t)$ will oscillate asymptotically.
This shows how the perturbative series (\ref{powerSol}) behaves
beyond its convergence radius, so that $\ugd(\kt)$ stays marginally
square-integrable,%
\footnote{That is, $\ugd(\kt)\sim(\kt^2)^{-1/2} \times \text{oscillating
    function}$.}
as required by the $\gamma$-representation.

Secondly, Eq.~(\ref{d:gluon}) shows, in the perturbative regime, a fluctuation
factor $\ff$ and an anomalous dimension exponential, in the form
\begin{gather}
 g_{\e}(t) = \ff\Big(\frac{\asb(t)}{\om}\Big)
 \exp\left\{\int_{-\infty}^t \dif\tau\;\left[
 \gamma_0\Big(\frac{\asb(\tau)}{\om}\Big)
 + \e \gamma_1\Big(\frac{\asb(\tau)}{\om}\Big) \right]\right\}
 \times\big[1+\ord{\e}\big]
\label{gForm} \\
 \ff\Big(\frac{\asb(t)}{\om}\Big) \equiv
 \frac1{\bar{\gamma}_t\sqrt{-\chi_0'(\bar{\gamma}_t)}} \;,
\label{d:ff}
\end{gather}
where
\begin{equation}\label{d:gamma1}
 \gamma_1\Big(\frac{\asb(t)}{\om}\Big) \equiv
 -\frac{\chi_1(\bar{\gamma}_t)}{\chi_0'(\bar{\gamma}_t)}
\end{equation}
is the $\ord{\e}$ correction to the BFKL anomalous dimension. This
interpretation follows simply by identifying the exponents in (\ref{gForm}) and
(\ref{d:gluon}), which have the same $t$-derivative (by the saddle-point condition
(\ref{saddle2})) and the same value at $t\to -\infty$ ($\bar{\gamma}_t=0$). Note
that the $1/\e$ singularity of the exponent is a consequence of the boundary at
$t\to -\infty$ in (\ref{gForm}), because of the Jacobians
\begin{equation}\label{jacobian}
 \dif t = \frac{\dif\asb}{\e\asb}
 = -\frac{\chi_0'(\bar{\gamma}_t)}{\e\chi_0(\bar{\gamma}_t)}\dif\bar{\gamma}_t \;,
\end{equation}
which relate the various forms of the $1/\e$ exponent:
\begin{equation}\label{exponents}
 \int_{-\infty}^t \dif\tau\; \gamma_0\Big(\frac{\asb(\tau)}{\om}\Big)
 = \frac1{\e}\int_0^{\asb(t)} \frac{\dif\alpha}{\alpha}\;
 \gamma_0\Big(\frac{\alpha}{\om}\Big) = \bar{\gamma}_t \, t
 +\frac1{\e}\int_0^{\bar{\gamma}_t} \dif\gamma'\; L_0(\gamma') \;.
\end{equation}

We are now in a position to factorise the $t$-dependence from the $1/\e$
singularities in Eq.~(\ref{gForm}). By simply subdividing the $\tau$ integration
interval into $]-\infty,0] \cup [0,t]$, we obtain %
\footnote{Splitting the integration interval at $\tau=0$ corresponds to choose
  the factorisation scale $\mu_f=\mu$, i.e., $t_f=0$.  }
\begin{equation}\label{gFact}
 g_{\e}(t) = \ff\Big(\frac{\asb(t)}{\om}\Big)
 \exp\left\{\int_0^t\dif\tau\;\gamma_0\Big(\frac{\asb(\tau)}{\om}\Big)
 \right\} \rk\Big(\frac{\asb(0)}{\om}\Big)
 \exp\left\{\frac1{\e}\int_0^{\asb(0)} \frac{\dif\alpha}{\alpha}\;
 \gamma_0\Big(\frac{\alpha}{\om}\Big) \right\} \;,
\end{equation}
where we have defined the reduced coefficient factor
\begin{equation}\label{d:rk}
 \rk(a) \equiv \frac{R(a)}{\ff(a)} \equiv \exp\left\{\int_0^{\gamma_0(a)}
   \dif\gamma'\;\frac{\chi_1(\gamma')}{\chi_0(\gamma')} \right\} \;, \qquad
 \left( a \equiv \frac{\asb}{\om} \right) \;,
\end{equation}
which arises because of the anomalous dimension correction $\e\gamma_1$
cancelling the $1/\e$ singularity of the Jacobian (\ref{jacobian}) according to the
identity
\begin{equation}\label{intGamma1}
 \e\int_{-\infty}^0 \dif\tau\;
 \gamma_1\Big(\frac{\asb(\tau)}{\om}\Big)
 = \int_0^{\asb(0)} \frac{\dif\alpha}{\alpha} \;
 \gamma_1\Big(\frac{\alpha}{\om}\Big)
 = \int_0^{\gamma_0\big(\frac{\asb(0)}{\om}\big)} \dif\gamma'\;
 \frac{\chi_1(\gamma')}{\chi_0(\gamma')} \;.
\end{equation}
The expression (\ref{d:rk}) agrees, by Eq.~(\ref{chi1}), with Eqs.~(3.17,B.18) of
Ref.~\cite{CaHa94}. Finally, the finite $t$-evolution in Eq.~(\ref{gFact})
becomes simply
\begin{equation}\label{tEvol}
 \ff\Big(\frac{\asb}{\om}\Big)\left(\frac{\kt^2}{\mu^2}\right)
 ^{\gamma_0\left(\frac{\asb}{\om}\right)}
\end{equation}
in the $\e\to0$ limit, thus recovering the result~\cite{CaHa94}
\begin{equation}\label{gRen}
 g_{\e}(t) \to R\Big(\frac{\asb}{\om}\Big)\left(\frac{\kt^2}{\mu^2}\right)
 ^{\gamma_0\left(\frac{\asb}{\om}\right)}
 \exp\left\{\frac1{\e}\int_0^{\asb} \frac{\dif\alpha}{\alpha}\;
 \gamma_0\Big(\frac{\alpha}{\om}\Big)
 \right\} \;.
\end{equation}
This derivation shows the mechanism by which the $R$ coefficient factor is
obtained as the product of $\ff$ (arising from the $\gamma$-fluctuations) and
$\rk$ (arising from the $\e$-dependence of the BFKL anomalous dimension). In the
following we shall generalise this mechanism to $b>0$ and to further terms in
the $\e$-expansion.

\section{The running coupling equation
  and its factorisation properties\label{s:rce}}

Our first purpose is to generalise to $4+2\e$ dimensions the BFKL equation with
running coupling. Because of the $\e$-dependence of the $\beta$-function in
Eq.~(\ref{beta}), the running coupling $\asb(t)$ has the form
\begin{equation}\label{runningB}
 \frac1{\asb(t)}-\frac{b}{\e}
 = \esp{-\e t} \left( \frac1{\asbmu}-\frac{b}{\e} \right)
 \;, \quad \text{or} \quad \asb(t)
 = \frac{\asbmu \esp{\e t}}{1+b\asbmu\frac{\esp{\e t}-1}{\e}} \;,
\end{equation}
where $b = 11/12$ in the $N_f = 0$ limit.
Due to the UV fixed point of Eq.~(\ref{beta}) at $\asb=\e/b$,
Eq.~(\ref{runningB}) shows two distinct regimes, according to whether {\it(i)}
$\asbmu < \e/b$ or {\it (ii)} $\asbmu > \e/b$. In the regime {\it (i)}
$\asb(t)$ runs monotonically from $\asb = 0$ to $\asb = \e/b$ for
$-\infty < t < +\infty$, while in the regime {\it (ii)} $\asb(t)$, starting from
$\e/b$ in the UV limit, goes through the Landau pole at
$t_\Lambda = \log(1-\e/b\asbmu)/\e < 0$, and reaches $\asb = \e/b$ from
below in the IR limit.

Since the LL$x$ kernel $\Kernel(t,t')\dif t'$ scales like $\esp{\e t}$, an
equation which realises such $\asb(t)$-evolution (and regimes) is obtained by
setting
\begin{align}
 \ugd_{\e,b}(\kt) &= \delta^{(2+2\e)}(\kt) + \frac1{\om} \,
 \frac1{1+b\asbmu\frac{\esp{\e t}-1}{\e}}
 \int\frac{\dif^{2+2\e}\kt'}{(2\pi)^{2+2\e}}\;\Kernel(\kt,\kt')\ugd_{\e,b}(\kt')
 \label{BFKLeqB} \\
 &= \delta^{(2+2\e)}(\kt) + \frac{\esp{\e\psi(1)}}{(\pi\kt^2)^{1+\e}}
 \widetilde{\ugd}_{\e,b}(\kt) \;.
\nonumber
\end{align}
It is soon realised that $\widetilde{\ugd}_{\e,b}(\kt)$, given by (cf.\ Eq.~(\ref{inteq}))
\begin{equation}\label{tildeEq}
 \widetilde{\ugd}_{\e,b}(\kt) = \frac1{\om}
 \frac{\asbmu \esp{\e t}}{1+b\asbmu\frac{\esp{\e t}-1}{\e}}
 \left[ 1 + K_{\e} \widetilde{\ugd}_{\e,b} \right] \;,
\end{equation}
has a well-defined iterative solution in the regime {\it (i)} in which
$0<\asb(t)<\e/b$. In fact, the $\kt$-integrations are convergent in the IR
because of $\asb(t)\sim\esp{\e t}$ for $t\to-\infty$, and everywhere else
because $\asb(t)$ is bounded. In this sense, $b,\e>0$ in the regime {\it (i)}
act as regulators of both the IR and UV regions, and of the Landau pole, so that
Eq.~(\ref{BFKLeqB}) is meaningful without any cutoffs --- a somewhat unique case
in BFKL theory. In the $b \to 0$ limit Eq.~(\ref{BFKLeqB}) reduces to
Eq.~(\ref{BFKLeq}), but become less convergent in the UV region: as noticed in
App.~\ref{aa:is}, only a finite number of iterations $n<1/\e$ is actually
possible for $b \neq 0$, for the UV integrations to be convergent.

Therefore we shall solve Eq.~(\ref{BFKLeqB}) in the regime {\it (i)}, where it
is nicely convergent, and we shall study the factorisation of $1/\e$
singularities by letting $b,\e \to 0$ with $\e/b > \asbmu$ kept
fixed. Eventually, we are instead interested in the physical limit of $\e \to 0$
with $\asbmu$ and $b$ kept fixed, but we shall perform it only at the end,
after factorisation of the $1/\e$ poles.

Let us introduce the $\gamma$-representation (\ref{d:gammaRep}) into
Eq.~(\ref{tildeEq}) after multiplying it by the $t$-dependent denominator. After
simple algebra we get the equation
\begin{align}
 f_{\e,b}(\gamma+\e)
 &= \frac1{\om}\left(\frac1{\asbmu}-\frac{b}{\e}\right)^{-1}
 \left[ \frac{\esp{\gamma T}}{\gamma} + \left( \chie(\gamma) -
 \frac{b\om}{\e} \right) f_{\e,b}(\gamma) \right]
\label{fdeB} \\ \nonumber
 &\equiv f_{\e,b}^{(0)}(\gamma) + \chi_{\e,b}(\gamma) f_{\e,b}(\gamma) \;,
\end{align}
where we have again introduced the cutoff $t > -T$ in the initial condition
(cf.\ Eq.~(\ref{F0})), in order to better control the $\e \to 0$
limit. Eq.~(\ref{fdeB}) has the form of Eq.~(\ref{fde}) and admits a similar
iterative solution, which is obtained from Eq.~(\ref{itSol}) by the replacements
\begin{subequations}\label{repls}
\begin{align}
 \asbmu &\to \left(\frac1{\asbmu}-\frac{b}{\e}\right)^{-1} \;,
\label{rep:alpha} \\
 f^{(0)}(\gamma) &\to f_{\e,b}^{(0)}(\gamma) \equiv \frac1{\om}
 \left(\frac1{\asbmu}-\frac{b}{\e}\right)^{-1}\frac{\esp{\gamma T}}{\gamma}\;,
\label{rep:f0} \\
 \chie(\gamma) &\to \chi_{\e,b}(\gamma) \equiv \chie(\gamma)-\frac{b\om}{\e} \;.
\label{rep:chi}
\end{align}
\end{subequations}
In the limit of $b,\e \to 0$ with $b/\e$ and the cutoff $T$ kept fixed we get
identically from Eqs.~(\ref{itSol}) and (\ref{repls}) the frozen coupling
solution (\ref{eps0lim}) and (\ref{frozenSol}), as expected.

We are more interested, however, in the continuum limit of the solution of
Eq.~(\ref{fdeB}) in the regime {\it (i)} and for $T \to +\infty$. By formal
manipulations similar to the $b=0$ case we obtain
\begin{equation}\label{gammaRepB}
 \widetilde{\ugd}_{\e,b}(\kt) \stackrel{T\to +\infty}{\longrightarrow}
 \int\frac{\dif\gamma}{\sqrt{2\pi\e}} \frac1{\sqrt{\chi_0(\gamma)-\frac{b\om}{\e}}}
 \exp\left\{ \gamma t+\frac1{\e}\int_0^{\gamma} L_{0,b}(\gamma')\dif\gamma'
 + \int_0^{\gamma} \frac{\chi_1(\gamma')}{\chi_0(\gamma')-\frac{b\om}{\e}}
 \dif\gamma' + \ord{\e} \right\}
\end{equation}
which is again a solution of the homogeneous equation, where
\begin{equation}\label{d:Lb}
 L_{\e,b}(\gamma) \equiv \log\left(
 \frac{\frac1{\om}\chie(\gamma) - \frac{b}{\e}}{\frac1{\asbmu}-\frac{b}{\e}}
 \right) \;, \quad
 L_{0,b}(\gamma) \equiv \log\left(
 \frac{\frac1{\om}\chi_0(\gamma) - \frac{b}{\e}}{\frac1{\asbmu}-\frac{b}{\e}}
 \right)
\end{equation}
and the exponent has been expanded in $\e$ (at $b \asb/\e$ fixed) up to the
finite terms. Once again, the factorisation of the $1/\e$ poles in
(\ref{gammaRepB}) is investigated, for $\e \to 0$, by the saddle point condition
\begin{equation}\label{saddleB}
 \e t + L_{0,b}(\bar{\gamma}) = 0 \quad \iff \quad
 1 = \frac{\asb(t)}{\om}\chi_0(\bar{\gamma})
\end{equation}
where, due to the $b$-dependence of Eq.~(\ref{d:Lb}), $\asb(t)$ has the
expected form (\ref{runningB}). It follows that
\begin{equation}\label{ugdTildeB}
 \widetilde{\ugd}_{\e,b}(\kt) = \frac1{\sqrt{-\chi_0'(\bar{\gamma}_t)}}
 \exp\left\{\bar{\gamma}_t\,t + \frac1{\e}\int_0^{\bar{\gamma}_t}
 L_{0,b}(\gamma')\dif\gamma' + \int_0^{\bar{\gamma}_t}
 \frac{\chi_1(\gamma')}{\chi_0(\gamma')-\frac{b\om}{\e}}\dif\gamma'
 \right\} \times \big[1+\ord{\e}\big]
\end{equation}
and that
\begin{align}
 g_{\e,b}(t) &= \frac1{\bar{\gamma}_t\sqrt{-\chi_0'(\bar{\gamma}_t)}}
 \exp\left\{\bar{\gamma}_t \,t + \frac1{\e}\int_0^{\bar{\gamma}_t}
 L_{0,b}(\gamma')\dif\gamma' + \int_0^{\bar{\gamma}_t}
 \frac{\chi_1(\gamma')}{\chi_0(\gamma')-\frac{b\om}{\e}}\dif\gamma'
 \right\} \times \big[1+\ord{\e}\big]
\nonumber \\
 &= \ff\Big(\frac{\asb(t)}{\om}\Big)
 \exp\left\{\int_{-\infty}^t \dif\tau\;\left[
 \gamma_0\Big(\frac{\asb(\tau)}{\om}\Big)
 + \e \gamma_1\Big(\frac{\asb(\tau)}{\om}\Big) \right]\right\}
 \times\big[1+\ord{\e}\big] \;.
\label{gFormB}
\end{align}

The final expression in Eq.~(\ref{gFormB}) is formally identical to
Eq.~(\ref{gForm}) and is expected to have analogous factorisation
properties. However, due to the different form of $\asb(t)$, the Jacobians
induced by Eq.~(\ref{saddleB}) are different:
\begin{equation}\label{jacobianB}
 \dif t = \frac{\dif\asb}{\asb(\e-b\asb)}
 = -\frac{\chi_0'(\bar{\gamma}_t)}{\e\left[\chi_0(\bar{\gamma}_t)
 -\frac{b\om}{\e}\right]} \dif\bar{\gamma}_t \;,
\end{equation}
and this explains the different $\e$-dependence of the exponents at fixed value
of $\asb(t)$:
\begin{equation}\label{exponent0B}
 \int_{-\infty}^t \dif\tau\; \gamma_0\Big(\frac{\asb(\tau)}{\om}\Big)
 = \int_0^{\asb(t)} \frac{\dif\alpha}{\alpha(\e-b\alpha)}\;
 \gamma_0\Big(\frac{\alpha}{\om}\Big)
 = \bar{\gamma}_t \, t
 +\frac1{\e}\int_0^{\bar{\gamma}_t} \dif\gamma'\; L_{0,b}(\gamma')
\end{equation}
and
\begin{equation}\label{exponent1B}
 \e\int_{-\infty}^t \dif\tau\; \gamma_1\Big(\frac{\asb(\tau)}{\om}\Big)
 = \e\int_0^{\asb(t)} \frac{\dif\alpha}{\alpha(\e-b\alpha)}\;
 \gamma_1\Big(\frac{\alpha}{\om}\Big) \;.
\end{equation}

The factorisation of the $t$-dependence in Eq.~(\ref{gFormB}) from the $1/\e$
singularities is now performed as in Sec.~\ref{s:grLL} by subdividing the
$\tau$-integration into the $]-\infty,t_f]$ and $[t_f,t]$ intervals, where
$t_f \equiv \log(\mu_f^2/\mu^2)$ defines the factorisation scale. The two
intervals are treated differently: in the finite UV interval $[t_f,t]$ we can
freely go to the $\e=0$ limit at fixed value of $\asb(t_f)$, thus recovering the
$\asb(t) \to \asb(t_f)/[1+b\asb(t_f) (t-t_f)]$ limit and the normal
$t$-dependence of UV free QCD in 4 dimensions. In the remaining infinite IR
interval we have to factorise the $1/\e$ poles by expanding in the $b\asb/\e$
parameter, as normally done in fixed order perturbation theory.

By looking at Eqs.~(\ref{exponent0B}) and (\ref{exponent1B}) (with $t$ replaced
by $t_f$) we realise the following: Eq.~(\ref{exponent0B}) exponentiates $1/\e$
poles and higher order ones coming from the $b\asb/\e$-expansion, and has to be
factorised in full in the $\MSbar$-scheme; it evolves in $t_f$ according to the
LL$x$ anomalous dimension.  On the other hand, the term (\ref{exponent1B}) ---
due to the $\ord{\e}$ correction to the BFKL anomalous dimension --- reduces, in
the $b \to 0$ limit, to the $\rk\big(\asb(t_f)/\om\big)$ factor found in
Sec.~\ref{s:grLL}, which yields $R/\ff$, i.e., the normalisation mismatch of the
frozen coupling evolution with respect to the $\MSbar$ density. Furthermore, in
the $b\asb/\e$-expansion, Eq.~(\ref{exponent1B}) exponentiates $1/\e$ and
higher order poles which should be factored out in the $\MSbar$ density, and
therefore contribute to the $\MSbar$ anomalous dimension.

In order to perform collectively such separation, we rewrite the r.h.s.\ of
Eq.~(\ref{exponent1B}) (with $t = t_f$) in the form
\begin{equation}\label{gamma1split}
 \int_0^{\asb(t_f)} \frac{\dif\alpha}{\alpha}\; \gamma_1\Big(\frac{\alpha}{\om}\Big) +
 \int_0^{\asb(t_f)} \frac{\dif\alpha}{\alpha(\e-b\alpha)}\; b\alpha \,
 \gamma_1\Big(\frac{\alpha}{\om}\Big) \;,
\end{equation}
where the first term provides $\log\rk\big(\asb(t_f)/\om\big)$, and the second
one is the anomalous dimension contribution to be factored out in the $\MSbar$
density. Finally, by replacing Eq.~(\ref{gamma1split}) into Eq.~(\ref{gFormB}),
we find
\begin{equation}\label{gQ0}
 g_{\e}(t) = \ff\Big(\frac{\asb(t)}{\om}\Big)
 \exp\left\{\int_{t_f}^t\dif\tau\;\gamma_0\Big(\frac{\asb(\tau)}{\om}\Big)
 \right\} \rk\Big(\frac{\asb(t_f)}{\om}\Big) g^{(\MSbar)}(t_f)\;,
\end{equation}
where
\begin{equation}\label{gMSbar}
 g^{(\MSbar)}(t_f) =
 \exp\left\{\int_0^{\asb(t_f)} \frac{\dif\alpha}{\alpha(\e-b\alpha)}\;
 \left[ \gamma_0\Big(\frac{\alpha}{\om}\Big)
 +b\alpha \gamma_1\Big(\frac{\alpha}{\om}\Big) \right] \right\}
\end{equation}
incorporates all the $1/\e$ poles, in a formal $b\asb/\e$-expansion.

Eqs.~(\ref{gQ0}) and (\ref{gMSbar}) are the main results of this section. They
confirm the relation between our generalised $Q_0$-scheme (with dimensional
regularisation) and the $\MSbar$-scheme, namely
\begin{equation}\label{schemeRel}
 g^{(Q_0)}(t) \equiv g_{\e}(t)
 = R\Big(\frac{\asb(t)}{\om}\Big) g^{(\MSbar)}(t) \;,
\end{equation}
and they prove the following resummation formulas
\begin{align}
 \gamma^{(Q_0)}\big(\asb(t);\om\big)
 \equiv \frac{\frac{\dif}{\dif t} g^{(Q_0)}(t)}{g^{(Q_0)}(t)}
 &= \gamma_0\Big(\frac{\asb(t)}{\om}\Big)
 -b \asb(t) \frac{\partial\log \ff}{\partial\log\asb}
\label{gammaQ0} \\
 \gamma^{(\MSbar)}\big(\asb(t);\om\big)
 \equiv \frac{\frac{\dif}{\dif t} g^{(\MSbar)}(t)}{g^{(\MSbar)}(t)}
 &= \gamma_0\Big(\frac{\asb(t)}{\om}\Big)
 +b \asb(t) \frac{\partial\log \rk}{\partial\log\asb}
\label{gammaMSbar} \\ \nonumber
 &= \gamma_0\Big(\frac{\asb(t)}{\om}\Big)
 +b \asb(t) \gamma_1\Big(\frac{\asb(t)}{\om}\Big)
\end{align}
where $\ff$, $\rk$, $\gamma_0$ and $\gamma_1$ are defined in Eqs.~(\ref{d:ff}),
(\ref{d:rk}), (\ref{d:gamma0}) and (\ref{d:gamma1}) respectively.

Let us remark that the results (\ref{gammaQ0}-\ref{gammaMSbar}) are not
surprising, in view of the well studied relationship of the $\ff$ and $R$
factors already performed in the literature in a somewhat different
context~\cite{CaCi97}. However they offer some important insights into the
explicit RG factorisation obtained here: Eq.~(\ref{gammaQ0}) shows that the
evolution of the gluon density in a $\kt$-factorisation scheme is independent of
whether the BFKL equation is regularised by a cutoff ($Q_0$-scheme in strict
sense) or by a positive $\e$ (as done at present). This is a consequence
of the RG factorisation just proved. Eq.~(\ref{gammaMSbar}) shows that the
$\MSbar$ evolution is the one expected because of the $R$ factor in
Eq.~(\ref{schemeRel}). However, this result is obtained by the explicit,
$b$-dependent factorisation of the $1/\e$ poles connected with $\gamma_1$ in
Eq.~(\ref{gMSbar}). Furthermore the $t_f$-dependence in Eq.~(\ref{gQ0}) cancels
out because of the $\gamma_1$ evolution in $g^{(\MSbar)}$ cancelling the
$t_f$-dependence of $\rk$. All together, such terms build up the $\ord{\e}$
correction to the BFKL anomalous dimension, as done in Eq.~(\ref{gamma1split}),
which precisely vanishes in the $\e \to 0$ limit.

Our next goal is to generalise the above factorisation procedure to subleading
terms, in particular to the coefficient terms at NL$x$ level and to the
corresponding NNL$x$ anomalous dimension terms.

\section{Factorisation at subleading level:
  $\boldsymbol{b}$-dependent corrections\label{s:fsl}}

We have just realised that the normalisation factor $\rk$ arises from the
$\ord{\e}$ correction to the BFKL anomalous dimension (\ref{exponent1B}) after
the (minimal) subtraction of a $b$-dependent contribution to the $\MSbar$
anomalous dimension (second term of (\ref{gamma1split})). This suggests that, in
order to compute corrections of relative order $b\asb$ to $\rk$ and the
corresponding ones to $\gamma^{(\MSbar)}$ we have to calculate $\ord{\e^2}$
corrections to the BFKL anomalous dimension and, more generally, to the exponent
of the solution (\ref{gammaRepB}) of the homogeneous equation. We thus restart
our analysis from the ``action'' (\ref{action}) and its $b/\e$-dependent
counterpart, and we rewrite the solution (\ref{gammaRepB}) of the homogeneous
equation in the form
\begin{equation}\label{solHomEqEps2}
 \widetilde{\ugd}_{\e,b}(\kt) =
 \int\frac{\dif\gamma}{\sqrt{2\pi\e}}
 \frac1{\sqrt{\chi_{\e}(\gamma)-\frac{b\om}{\e}}} \exp\left\{
 \gamma t+\frac1{\e}\int_0^{\gamma} L_{\e,b}(\gamma')\dif\gamma'
 + \frac{\e}{12} L_{\e,b}'(\gamma) + \ord{\e^2} \right\} \;,
\end{equation}
where we have kept the $\e$-dependence of $L$ and $\chi$, and the $\ord{\e}$
correction to the ``action'' (which is of relative order $\ord{\e^2}$).

We then look at the factorisation properties of Eq.~(\ref{solHomEqEps2}) by
expanding around the $\e$-dependent saddle point $\gamma=\bar{\gamma}_{\e}$:
\begin{equation}\label{saddleEps2}
 \e t + L_{\e,b}(\bar{\gamma}_{\e}) = 0 \quad \iff \quad 1
 = \frac{\asb(t)}{\om} \chie(\bar{\gamma}_{\e}) \;,
\end{equation}
which, therefore, defines the $\e$-dependent BFKL anomalous dimension
\begin{equation}\label{gammaEps2}
 \gamma_{\e}\Big(\frac{\asb}{\om}\Big)
 = \gamma_0\Big(\frac{\asb}{\om}\Big) + \e \gamma_1\Big(\frac{\asb}{\om}\Big)
 + \e^2 \gamma_2\Big(\frac{\asb}{\om}\Big) + \cdots
\end{equation}
where $\gamma_1$ was given before in Eq.~(\ref{d:gamma1}), and we have
\begin{equation}\label{gamma2}
 \gamma_2\Big(\frac{\asb}{\om}\Big)
 = \left. -\frac{\chi_2(\gamma)}{\chi_0'(\gamma)}
 + \frac{\chi_1(\gamma)
 \chi_1'(\gamma)}{\chi_0'{}^2(\gamma)}
 -\frac12\frac{\chi_1^2(\gamma) \chi_0''(\gamma)}
 {\chi_0'{}^3(\gamma)} \right|_{\gamma = \gamma_0(\frac{\asb}{\om})} \;.
\end{equation}

Furthermore, we have to compute the $\gamma$-integral by expanding the action
and the factor $\left[\chie(\gamma)-\frac{b\om}{\e}\right]^{-1/2}$
around the saddle point. Since the $\gamma$-fluctuations are governed by the
width $\sigma_\gamma = \left[-\e/L'_{\e,b}(\bar{\gamma}_{\e})\right]^{1/2}$
which is of $\ord{\sqrt{\e}}$ while the ``action'' is $\ord{1/\e}$, we need to
expand, it turns out, up to 6-th order in $\gamma-\bar{\gamma}$ in order to
reach all $\ord{\e}$ terms. This calculation is performed in App.~\ref{a:cgd} and,
when replaced into Eq.~(\ref{solHomEqEps2}), provides the following result for
the gluon density:
\begin{equation}\label{gluonEps2}
 g_{\e}(t) = \frac1{\bar{\gamma}_{\e}\sqrt{-\chie'(\bar{\gamma}_{\e})}}
 \exp\left\{ \int_{-\infty}^t \dif\tau\;
 \bar{\gamma}_{\e}\Big(\frac{\asb(\tau)}{\om}\Big) \right\} \times
 \big[ 1 + \e S_1(\bar{\gamma}_{\e},b) \big] \;.
\end{equation}
Here we have used the identity
\begin{equation}\label{expintgamma}
 \bar{\gamma}_{\e}\, t +\frac1{\e}\int_0^{\bar{\gamma}_{\e}}\dif\gamma'\;
 L_{\e,b}(\gamma') = \int_{-\infty}^t \dif\tau\;
 \bar{\gamma}_{\e}\Big(\frac{\asb(\tau)}{\om}\Big) \;,
\end{equation}
the Jacobians
\begin{equation}\label{jacobEps}
 \dif t = \frac{\dif\asb}{\asb(\e-b\asb)} =
 -\frac{\chie'(\bar{\gamma}_{\e})}{\e\left[\chie(\bar{\gamma}_{\e})
 -\frac{b\om}{\e}\right]} \dif \bar{\gamma}_{\e} \;,
\end{equation}
and the $\ord{\e}$ correction to the action
\begin{equation}\label{S1}
 S_1(\gamma,b) = \frac1{12} L_{\e,b}'+\left[\frac1{8}(-L_{\e,b}') + \frac{5}{24}
 \frac{L_{\e,b}''{}^2}{(-L_{\e,b}')^3}
 + \frac1{8}\frac{L_{\e,b}'''}{L_{\e,b}'{}^2} \right] \;,
\end{equation}
where the terms in square brackets are precisely the fluctuations calculated in
App.~\ref{aa:cspf}.

Let us now look at the factorisation properties of our result in
Eq.~(\ref{gluonEps2}) in the limit of $\e, b \asb/\e \to 0$.
The anomalous dimension exponential is factorised as usual, and its infrared
part at the factorisation scale $t_f$ is given by
\begin{equation}\label{IRexponential}
 \int_0^{\asb(t_f)}\frac{\dif\alpha}{\alpha(\e-b\alpha)} \left[
 \gamma_0\Big(\frac{\alpha}{\om}\Big) + \e \gamma_1\Big(\frac{\alpha}{\om}\Big)
 + \e^2 \gamma_2\Big(\frac{\alpha}{\om}\Big) + \cdots \right] \;.
\end{equation}
We have now the additional $\ord{\e^2}$ term $\gamma_2$, which however still
builds $1/\e$ singularities because of the $b\asb/\e$-expansion of the
$\beta$-function in the denominator. The decomposition into coefficient and
$\MSbar$-anomalous dimension parts is simply done by writing
$\e^2 = (b\asb)^2 + (\e-b\asb)(b\asb+\e)$, so that
\begin{equation}\label{gamma2split}
 \int_0^{\asb(t_f)}\frac{\dif\alpha}{\alpha(\e-b\alpha)} \;
 \e^2 \, \gamma_2\Big(\frac{\alpha}{\om}\Big) =
 \int_0^{\asb(t_f)}\frac{\dif\alpha}{\alpha(\e-b\alpha)} \;
 (b\alpha)^2 \, \gamma_2\Big(\frac{\alpha}{\om}\Big) +
 \int_0^{\asb(t_f)}\frac{\dif\alpha}{\alpha}\;
 b\alpha \,\gamma_2\Big(\frac{\alpha}{\om}\Big) +\ord{\e} \;,
\end{equation}
where the first term in the r.h.s.\ is the minimal subtraction part of the
$\gamma_2$ anomalous dimension, the second one is the $\ord{b\asb}$ correction to
$\log\rk$.

We have also to look for $1/\e$ singularities possibly arising because of the
$b\asb/\e$-expansion in the contribution (\ref{S1}). Here the situation is not
so clear {\em a priori}, given the fact that
\begin{equation}\label{Lprime}
 -L_{\e,b}'(\bar{\gamma}_{\e}) =
 \frac{-\chie'(\bar{\gamma}_{\e})}{\chie(\bar{\gamma}_{\e})-\frac{b\om}{\e}}
 = -\frac{\asb(t)}{\om} \frac{\chie'\big(\bar{\gamma}_{\e}(t)\big)}
 {1-\frac{b \asb(t)}{\e}}
\end{equation}
has a non-trivial $b\asb/\e$ expansion. However, after some algebra, we find a
remarkable cancellation leading to the result (see App.~\ref{aa:cs})
\begin{equation}\label{S1form}
 S_1\big(\bar{\gamma}_{\e}(t),b\big) = \frac1{24} \left\{ \frac{\chie''}{\chie'}
 + \frac{\om}{\chie'}\left(\frac{b}{\e}-\frac1{\asb(t)}\right)
 \left[2\left(\frac{\chie''}{\chie'}\right)^2
 -3\left(\frac{\chie''}{\chie'}\right)'\right]\right\} \;,
\end{equation}
which is only linear in $b/\e$. It follows that $S_1$ has no minimal subtraction
terms, and only contributes to the renormalisation of the $\ff$-factor, as
follows:
\begin{equation}\label{epsS1}
 \e S_1\big(\bar{\gamma}_{\e}(t),b\big) = \frac{b\om}{24\chi_0'}\left[
 2\left(\frac{\chi_0''}{\chi_0'}\right)^2
 -3\left(\frac{\chi_0''}{\chi_0'}\right)'\right] + \ord{\e} \;.
\end{equation}

We can thus summarise our results on the NL$x$ corrections to the coefficient
factors of Eq.~(\ref{gQ0}) (and on the NNL$x$ ones to the $\MSbar$ anomalous
dimension). The $\rk$ factor takes the form
\begin{equation}\label{rkEps2}
 \rk\big(\asb(t_f);\om\big) = \exp\left\{\int_0^{\asb(t_f)}
 \frac{\dif\alpha}{\alpha} \;\left[ \gamma_1\Big(\frac{\alpha}{\om}\Big)
 + b\alpha \, \gamma_2\Big(\frac{\alpha}{\om}\Big) + \cdots \right] \right\} \;,
\end{equation}
and the $\MSbar$ anomalous dimension becomes
\begin{align}
 \gamma^{(\MSbar)}(\asb;\om) &=
 \gamma_0\Big(\frac{\asb}{\om}\Big)
 +b \asb \, \gamma_1\Big(\frac{\asb}{\om}\Big)
 +b^2 \asb^2(t) \, \gamma_2\Big(\frac{\asb}{\om}\Big) +\cdots
\label{gammaMSbar2} \\ \nonumber
 &= \gamma_0\Big(\frac{\asb}{\om}\Big) + b \asb
 \frac{\partial\rk(\asb;\om)}{\partial\log\asb} \;.
\end{align}
Due to the form of (\ref{gammaMSbar2}) and (\ref{rkEps2}), the $t_f$-dependence
of the factorisation formula (\ref{gQ0}) cancels out as it should, and as is
expected on the basis of their common origin, Eq.~(\ref{IRexponential}). We
also expect this mechanism to hold true to all orders in $b\asb$, so that the
$\MSbar$ anomalous dimension and coefficient can be inferred from the
corresponding $\e$-expansion of the BFKL anomalous dimension (\ref{gammaEps2}).

On the other hand, the fluctuation factor $\ff$ takes finite NL$x$ corrections
from Eq.~(\ref{epsS1}), and becomes
\begin{equation}\label{ffEps2}
 \ff(\asb;\om) = \frac1{\gamma_0\sqrt{-\chi_0'}} \left\{ 1 +
 \frac{b\om}{24\chi_0'}\left[ 2\left(\frac{\chi_0''}{\chi_0'}\right)^2
 -3\left(\frac{\chi_0''}{\chi_0'}\right)'\right] +\cdots \right\} \;.
\end{equation}
Such corrections are identical to those found from the normal
$\gamma$-representation in 4 dimensions. In this case also we expect that higher
orders in the $\e$-expansion (\ref{action}) and the corresponding fluctuations
will combine so as to provide further finite subleading corrections to
Eq.~(\ref{ffEps2}). We have no formal proof of this expectation, but we notice
that higher order fluctuations still involve the scale $t$ and the coupling
$\asb(t)$ and are thus independent of the IR behaviour at $\tau \to -\infty$. It
is then natural to believe that they can be computed directly in the $\e = 0$
limit from the normal $\gamma$-representation in 4 dimensions, as explicitly
proved above for the next-to-leading terms.

\section{On the treatment of the full NL$\boldsymbol{x}$ corrections to the
  $\boldsymbol{\rk}$ factor\label{s:nlc}}

Having calculated the running coupling corrections to $\rk$, the problem arises
of including {\em all} NL$x$ contributions, as embodied in the next-to-leading
BFKL kernel. The structure and the explicit expression of such a kernel with
dimensional regularisation were given in several papers for the
gluon~\cite{RGvert} and the quark~\cite{QQvertCC,QQvertFFFK} parts, and
summarised in Refs.~\cite{CaCi97,FaLi98,CaCi98}. It has the form
\begin{equation}\label{NLstruc}
  \Kernel^{(\mathrm{NL})} = \asb \left[ \frac{b}{\e} \left(1-\esp{\e t}\right)
  \Kernel \right] + \Kernel^{(1)} \;,
\end{equation}
where we have singled out the running coupling part (proportional to the leading
kernel $\Kernel$) which arises from the $\ord{\asb^2}$ expansion of the running
coupling equation~(\ref{BFKLeqB}). The remaining kernel $\Kernel^{(1)}$, which
scales as $\esp{2\e t}$, is the properly called NL$x$ kernel, whose eigenvalues
have been worked out in the literature in the $\e \to 0$ limit, and are here
required up to $\ord{\e}$ for the complete calculation of NL$x$ corrections to
$\rk$.

We shall thus generalise Eq.~(\ref{BFKLeqB}) to include the NL$x$ kernel in the
form
\begin{align}
 \ugd_{\e}(\kt) &= \delta^{(2+2\e)}(\kt) + \frac1{\om} \,
 \frac1{1+b\asbmu\frac{\esp{\e t}-1}{\e}}
 \int\frac{\dif^{2+2\e}\kt'}{(2\pi)^{2+2\e}}\;\left[ \Kernel(\kt,\kt')
 + \Kernel^{(1)}(\kt,\kt') \right] \ugd_{\e}(\kt')
\nonumber \\
 &= \delta^{(2+2\e)}(\kt) + \frac{\esp{\e\psi(1)}}{(\pi\kt^2)^{1+\e}}
 \widetilde{\ugd}_{\e}(\kt) \;.
\label{BFKLeqNL}
\end{align}
where NNL$x$ terms and further subleading ones have been freely added so as to
reproduce the resummed $\asb(t)$ evolution.%
\footnote{Here we keep terms which are of relative order $\asb \e^n$ ($n \geq
  0$) with respect to the leading ones, and we drop terms of relative order
  $b\asb^2$ or higher, at fixed values of $\e$.}
It is now straightforward to go to $\gamma$-space and to obtain a modified form
of Eq.~(\ref{fdeB}). By introducing the NL$x$ ``characteristic function''
$\chie^{(1)}$ of $\Kernel^{(1)}$ in $4+2\e$ dimensions
\begin{equation}\label{d:chie1}
 \int\frac{\dif^{2+2\e}\kt'}{(2\pi)^{2+2\e}}\;
 \Kernel^{(1)}(\kt,\kt') (\kt'{}^2)^{\gamma-1-2\e}
 \equiv \asbmu^2 \chie^{(1)}(\gamma) \frac{(\kt^2)^{\gamma-1}}{\mu^{4\e}} \;,
\end{equation}
the corresponding homogeneous equation reads
\begin{equation}\label{fdeNL}
  f_{\e}(\gamma+\e) - \frac{b\asbmu}{\e}
  \left[ f_{\e}(\gamma+\e)-f_{\e}(\gamma)\right]
  = \asbmu \frac{\chie(\gamma)}{\om} f_{\e}(\gamma)
  + \frac{\asbmu^2}{\om}
  \chie^{(1)}(\gamma) f_{\e}(\gamma-\e) \;,
\end{equation}
and contains, therefore, two finite difference steps, due to the different
scaling properties of the leading vs.\ next-to-leading kernels. However, to
NL$x$ accuracy, we can replace the leading order equation in the last term, and
we obtain
\begin{equation}\label{fdeNLbis}
  f_{\e}(\gamma+\e) = \left[
  \frac{
    \frac{\chie(\gamma)}{\om}-\frac{b}{\e}
    + \frac{\chie^{(1)}(\gamma)}{\chie(\gamma-\e)}
  }
  {\frac1{\asbmu}-\frac{b}{\e}}
  \right] f_{\e}(\gamma)
  \equiv \exp[L_{\e,b}^{\mathrm{eff}}(\gamma)] f_{\e}(\gamma)
 \;,
\end{equation}
where the next-to-leading term is now suppressed by a factor of $\om$ with
respect to the leading one, much in the same spirit as the
$\om$-expansion~\cite{omExp}.

We can thus solve Eq.~(\ref{fdeNLbis}) by the same method used before to get
Eq.~(\ref{solHomEqEps2}).
Correspondingly, the saddle point at
$\gamma = \bar{\gamma}_{\e}\big(\asb(t);\om\big)$ such that
\begin{equation}\label{saddleNL}
  \e t + L_{\e,b}^{\mathrm{eff}}\left(\bar{\gamma}_{\e}\right) = 0
  \qquad \iff \qquad
  \asb(t) \left[ \chie(\bar{\gamma}_{\e}) + \om
  \frac{\chie^{(1)}(\bar{\gamma}_{\e})}{\chie(\bar{\gamma}_{\e}-\e)}
  \right] = \om
\end{equation}
admits the solution
\begin{equation}\label{gammaNL}
  \bar{\gamma}_{\e} = \gamma_{\e}^{(0)}\Big(\frac{\asb(t)}{\om}\Big)
  + \asb(t) \gamma_{\e}^{(1)}\Big(\frac{\asb(t)}{\om}\Big) + \ord{\asb^2} \;,
\end{equation}
where $\gamma_{\e}^{(0)}$ defines the LL$x$ BFKL anomalous dimension of
Eq.~(\ref{gammaEps2})
and $\gamma_{\e}^{(1)}$ is its NL$x$ correction, obtained by truncating the
expansion of the saddle point position to relative order $\ord{\asb(t)}$:
\begin{equation}\label{delta}
  \gamma_{\e}^{(1)}\Big(\frac{\asb(t)}{\om}\Big) = -
  \frac{\chie^{(1)}(\gamma_{\e}^{(0)})}{\chie'(\gamma_{\e}^{(0)})}
  \frac{\chie(\gamma_{\e}^{(0)})}{\chie(\gamma_{\e}^{(0)} - \e)}
  = \gamma^{(1)}_0\Big(\frac{\asb(t)}{\om}\Big)
  + \e \gamma^{(1)}_1\Big(\frac{\asb(t)}{\om}\Big) + \ord{\e^2} \;.
\end{equation}
This expression reduces to the customary one~\cite{FaLi98,CaCi98} in the $\e=0$
limit, but has $\ord{\e}$ corrections coming from the corresponding ones of
$\chie^{(1)}$ --- yet to be extracted from the various papers in the
literature~\cite{RGvert,QQvertCC,QQvertFFFK}.

Finally, we expand the $\e$-dependence of the BFKL anomalous dimension,
including NL$x$ terms, as follows:
\begin{equation}\label{gammaBFKL}
  \bar{\gamma}_{\e}(\asb;\om) = \gamma^{(0)}_0\Big(\frac{\asb}{\om}\Big)
  + \asb \gamma^{(1)}_0\Big(\frac{\asb}{\om}\Big)
  + \e \gamma^{(0)}_1\Big(\frac{\asb}{\om}\Big) + \e \left[
  \asb \gamma^{(1)}_1\Big(\frac{\asb}{\om}\Big)
  + \e \gamma^{(0)}_2\Big(\frac{\asb}{\om}\Big) \right]
\end{equation}
and we replace it into the analogue of Eq.~(\ref{gluonEps2}). The corresponding
anomalous dimension exponential factorises in the form
\begin{equation}\label{anDimExp}
  \exp\left\{\int_0^{\asb(t)} \frac{\dif\alpha}{\alpha(\e-b\alpha)} \;
  \bar{\gamma}_{\e}(\alpha;\om) \right\}
  = \rk\big(\asb(t);\om\big) g^{(\MSbar)}(t) \;,
\end{equation}
where
\begin{equation}\label{logrk}
 \rk\big(\asb(t);\om\big) = \exp \left\{ \int_0^{\asb(t)}
 \frac{\dif\alpha}{\alpha}\;
 \left[\gamma^{(0)}_1\Big(\frac{\alpha}{\om}\Big)
 + \alpha \gamma^{(1)}_1\Big(\frac{\alpha}{\om}\Big)
 + b \alpha \gamma^{(0)}_2\Big(\frac{\alpha}{\om}\Big) \right] \right\}
\end{equation}
and
\begin{equation}\label{gMSbarNL}
  g^{(\MSbar)}(t) = \exp\left\{\int_0^{\asb(t)}
 \frac{\dif\alpha}{\alpha(\e-b\alpha)} \;
  \bar{\gamma}_{\e=b\alpha}(\alpha;\om) \right\} \;.
\end{equation}

It follows that the $\MSbar$ gluon anomalous dimension is
\begin{align}
  \gamma^{(\MSbar)}(\asb;\om) = \bar{\gamma}_{\e=b\asb}(\asb;\om)
  &= \gamma^{(0)}_0 + \asb \gamma^{(1)}_0 + b\asb \gamma^{(0)}_1
  + b\asb^2 \gamma^{(1)}_1 + (b\asb)^2 \gamma^{(0)}_2
 \nonumber \\
  &= \gamma^{(0)}_0 + \asb\gamma^{(1)}_0
  + b\asb \frac{\partial\rk(\asb;\om)}{\partial\log\asb}
\label{gammaMSbarNL}
\end{align}
and has therefore subleading terms determined by the $\e$-dependence of
$\bar{\gamma}_{\e}$ at $\e = b\asb$, which are related in the expected way to
the coefficient $\rk$.

On the other hand, the gluon density in the $Q_0$-scheme contains an additional
fluctuation factor $\ff(\asb;\om)$ whose calculation at full NL$x$ level proceeds
along the lines of Sec.~\ref{s:fsl} and is not explicitly done here. The
corresponding anomalous dimension is
\begin{equation}\label{gammaQ0NL}
  \gamma^{(Q_0)}(\asb;\om) = \gamma^{(0)}_0\Big(\frac{\asb}{\om}\Big)
  + \asb \gamma^{(1)}_0\Big(\frac{\asb}{\om}\Big)
  -b \asb \frac{\partial \ff(\asb;\om)}{\partial\log\asb}
\end{equation}
and can be calculated directly at $\e = 0$.

It should be remarked that the NNL$x$ contributions to the anomalous dimension
in Eqs.~(\ref{gammaMSbarNL}) and (\ref{gammaQ0NL}) coming from the NL$x$
corrections to $\rk$ and $\ff$ are of course mixed with dynamical contributions
whose calculation has not yet been attempted in the literature. Nevertheless, we
do compute here --- once $\gamma_{\e}^{(1)}$ is known --- the full NL$x$
contributions to $\rk$, which therefore explain the {\em difference} between
$\MSbar$- and $Q_0$-scheme anomalous dimensions at NNL$x$ accuracy.

\section{Universality of $\boldsymbol{\gamma_{\pq\pg}^{(\MSbar)}}$ and its
  resummation formula\label{s:ugqg}}

\begin{figure}[hbp]
  \centering
  \includegraphics[height=0.2\textheight]{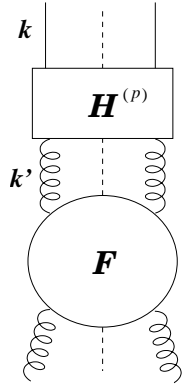}
  \caption{The leading small-$x$ contribution to the quark density.}
  \label{fig:quarkLx}
\end{figure}

So far, we have considered the gluon channel only, and discussed NL$x$
corrections to the normalisation change of the gluon density from the
$\MSbar$-scheme to the $Q_0$-scheme. However physical probes are coupled to
quarks, which enter the BFKL framework through a $\kt$-factorisation kernel
$\qKernel$, defining the measuring process at hand (labelled by the
superscript $p$), as shown in Fig.~\ref{fig:quarkLx}.

In a physical scheme, like the $Q_0$-scheme, we can just {\em define} the quark
density $q^{(p)}$ by the action of $\qKernel$
\begin{equation}
  \label{defQuark}
  q_{\e}^{(p)}(t) = \as(t) \int\dif^{2+2\e} \kt' \;
  \qKernel\Big(\frac{\kt^2}{\kt'{}^2}\Big) \ugd_{\e}(\kt') \;,
\end{equation}
($\as(t) \equiv \as \esp{\e t}$, see Eq.~(\ref{d:alphas}))
and it is then pretty easy to find a 4-dimensional NL$x$ resummation formula for
$\gamma_{\pq\pg}^{(p)}$, as first shown by Catani and Hautmann~\cite{CaHa94} in
the DIS-scheme ($q^{(\DIS)} = F_2$).

On the other hand, in the $\MSbar$-scheme one has to disentangle the coefficient
part of Eq.~(\ref{defQuark}) from the anomalous dimension part. The latter
contains, by definition, the minimal $1/\e$ singularities which are not
exponentiated in the gluon density in the following collinear factorisation
formula
\begin{equation}\label{cff}
  q_{\e}^{(p)}(t) - C_{\pq\pg}^{(p)}\big(\as(t)\big) g_{\e}^{(\MSbar)}(t)
  = q_{\e}^{(\MSbar)}(t)
  = \int_{-\infty}^t \dif\tau\; \gamma_{\pq\pg}^{(\MSbar)}
 \big(\as(\tau)\big) g_{\e}^{(\MSbar)}(\tau) \;,
\end{equation}
where we have set $C_{\pq\pq}^{(p)} = 1$ and $\gamma_{\pq\pq}^{(\MSbar)} = 0$ at
NL$x$ anomalous dimension accuracy and we have omitted, for notational
simplicity, the $\om$-variable. Note that $C_{\pq\pg}^{(p)}$ is process- and
$\e$-dependent, while $\gamma_{\pq\pg}$ is universal and $\e$-independent.

$q^{(p)}$ is obtained in the $\gamma$-representation framework by replacing in
Eq.~(\ref{defQuark}) the characteristic function of the $\qKernel$ kernel
\begin{equation}\label{defQuarkEigenv}
  \int \frac{\dif\kt'{}^2}{\kt'{}^2} \left(\frac{\kt'{}^2}{\kt^2}\right)^\gamma
  \qKernel\Big(\frac{\kt^2}{\kt'{}^2}\Big)
  \equiv \frac{ \qcf(\gamma) }{ \gamma (\gamma+\e) } \;,
\end{equation}
where the two poles are expected because of the intermediate
$\lt$-integration in $\qKernel$ ($\kt\gtrsim\lt\gtrsim\kt'$ roughly)
and the $\e$-dimension of the kernel integration providing also a running
$\as(t)=\as \cdot (\kt^2/\mu^2)^\e$
\begin{equation}
  \label{poleOrigin}
  \int\frac{\dif \kt'{}^2}{\kt'{}^2} \left(\frac{\kt'{}^2}{\kt^2}\right)^\gamma
  \int\frac{\dif^{2+2\e} \lt}{\lt^2} \; \Theta(\kt^2-\lt^2) \Theta(\lt^2-\kt'{}^2)
  =  c_\e \, \frac{(\kt^2)^{\e}}{\gamma(\gamma+\e)} \;.
\end{equation}
The pole $1/\gamma$ in (\ref{defQuarkEigenv}) is incorporated into the
integrated gluon density $g_{\e}(\gamma) = f_{\e}(\gamma)/\gamma$ to yield
\begin{equation}\label{quarkGammaRep}
  q_{\e}^{(p)}(t) = \as(t) \int \dif\gamma\; \esp{\gamma t} g_{\e}(\gamma)
  \frac{\qcf(\gamma)}{\gamma+\e} \;,
\end{equation}
where
\begin{equation}\label{gluonGammaRep}
  g_{\e}(t) = \int\dif\gamma\; \esp{\gamma t} g_{\e}(\gamma)
  = R\Big(\frac{\asb(t)}{\om},\e\Big) g_{\e}^{(\MSbar)}(t) \;.
\end{equation}

One could try a saddle point evaluation of $q$ around
$\gamma_0\big(\asb(t)/\om)\big)$, but it is difficult to get the right accuracy,
and the fluctuation algorithm is cumbersome. We prefer to extract
$\gamma_{\pq\pg}$ directly from the expression
\begin{subequations}\label{qdot}
\begin{align}
  \dot{q}_{\e}^{(p)}(t) \equiv \frac{\dif}{\dif t} q_{\e}^{(p)}(t)
  &= \as(t) \int\dif\gamma\; \esp{\gamma t} \, \qcf(\gamma) g_{\e}(\gamma)
 \label{qdot1} \\
  &= \frac{\dif}{\dif t} \left[ C_{\pq\pg}^{(p)}\big(\as(t),\e\big)
  g_{\e}^{(\MSbar)}(t) \right] + \gamma_{\pq\pg}\big(\as(t)\big)
  g_{\e}^{(\MSbar)}(t)
\label{qdot2} \\
  &= \left\{ \left[ \e \frac{\partial}{\partial\log\as(t)} +
  \gamma_0\Big(\frac{\asb(t)}{\om}\Big) \right]
  C_{\pq\pg}^{(p)}\big(\as(t),\e\big) + \gamma_{\pq\pg}^{(\MSbar)}
  \big(\as(t)\big) \right\} g_{\e}^{(\MSbar)}(t) \;,
 \label{qdot3}
\end{align}
\end{subequations}
having used in the first line the equality
\begin{equation}\label{gpe}
 \frac{\dif}{\dif t} \left[\as(t)\esp{\gamma t}\right]
  = \as(t)\esp{\gamma t} (\gamma+\e) \;.
\end{equation}

It is not obvious how to extract from Eq.~(\ref{qdot}) resummation formulas for
both $C_{\pq\pg}^{(p)}$ and $\gamma_{\pq\pg}$ in terms of the
$\kt$-factorisation integral of $\dot{q}_{\e}^{(p)}$. For instance, at $\e=0$ we
get the well known~\cite{CaHa94} $\kt$-factorisation result
\begin{align}
 \qRes_{0}^{(p)}\Big(\gamma_0\Big(\frac{\asb(t)}{\om}\Big)\Big)
 R\Big(\frac{\asb(t)}{\om},0\Big)
 &= C_{\pq\pg}^{(p)}\big(\as(t),0\big) \gamma_0\Big(\frac{\asb(t)}{\om}\Big)
 + \gamma_{\pq\pg}^{(\MSbar)}\big(\as(t)\big)
\nonumber \\
 &= \gamma_{\pq\pg}^{(p)}\big(\as(t)\big)
 R\Big(\frac{\asb(t)}{\om},0\Big)\;,
\label{qgFactEps0}
\end{align}
which determines $\gamma_{\pq\pg}$ in the $p$-scheme, but not $C_{\pq\pg}^{(p)}$
and $\gamma_{\pq\pg}^{(\MSbar)}$ separately. In Ref.~\cite{CaHa94}, a double
expansion of~(\ref{qdot2}) in both $\as(t)$ and $\e$ is used to get half a
dozen terms in the $\as(t)$-expansion of $\gamma_{\pq\pg}$, a result further
improved in~\cite{Rterms}.

Here we want to derive $\gamma_{\pq\pg}^{(\MSbar)}\big(\as(t)\big)$ to all
orders in $\asb(t)/\om$, by exploiting the property of the coefficient part
in~(\ref{qdot2}) of being a total $t$-derivative of the product of $g_{\e}$ and
a function which is perturbative in both $\as(t)$ and $\e$. We start noticing
that $\qcf$ can be expanded around $\gamma=-\e$ in the form
\begin{align}
  \qcf(\gamma) &= \qcf(-\e) + \ord{\gamma+\e}
\nonumber \\
  &\equiv \qRes(\e) + \ord{\gamma+\e} \;,
\label{qKerExpns}
\end{align}
where $\qRes(\e)$ is now process-independent, being the residue of the
function~(\ref{defQuark}) at the collinear pole $\gamma=-\e$. In
App.~\ref{a:cfqk} we evaluate $\qcf(\gamma)$ in various cases and we show that
in all cases $\qRes(\e)$ is process-independent and can be written as the
product of a rational and a transcendental part
\begin{equation}\label{qKernelEps}
 \qRes(\e) = \left[ \frac{T_R}{2\pi} \, \frac{2}{3} \,
 \frac{1+\e}{(1+2\e)(1+\frac{2}{3}\e)} \right]
 \left[ \frac{\esp{\e\psi(1)}\Gamma^2(1+\e)\Gamma(1-\e)}{\Gamma(1+2\e)} \right]
 \equiv \qResR(\e) \qResT(\e) \;.
\end{equation}
This form basically follows from the off-shell
generalisation of the $\pg\to\pq\bar{\pq}$ DGLAP splitting function introduced
in~\cite{CaHa94}, and proved in App.~\ref{aa:ucb} to have a universal off-shell
dependence induced by $\kt$-factorisation.

On the other hand, by Eq.~(\ref{gpe}) terms of order $(\gamma+\e)^n \;(n\geq1)$
are easily seen to be total $t$-derivatives, so that we can rewrite
Eq.~(\ref{qdot}) in the form
\begin{equation}\label{gqgxgMSbar}
  \qRes(\e) \as(t) R\Big(\frac{\asb(t)}{\om},\e\Big) g_{\e}^{(\MSbar)}(t)
  = \gamma_{\pq\pg}^{(\MSbar)}\big(\as(t)\big) g_{\e}^{(\MSbar)}(t) +
  \text{total $t$-derivative} \;.
\end{equation}
This result shows that $\gamma_{\pq\pg}^{(\MSbar)}$ is universal, being
dependent on the universal functions $\qRes(\e)$ and $R$. Furthermore, it
suggests how to extract the anomalous dimension contributions from the power
series in $\e$ in the l.h.s., i.e., by subtracting a total $t$-derivative or, in
other words, by doing an ``integration by parts''. By writing the expansions
\begin{align}
  \qResT(\e) R\Big(\frac{\asb(t)}{\om},\e\Big)
  &= \sum_{n=0}^{\infty} (-\e)^n R_n\Big(\frac{\asb(t)}{\om}\Big)
\label{Rexpns} \\
  \qResR(\e) &= \sum_{m=0}^{\infty} (-\e)^m \qRes_m
\label{qKexpns}
\end{align}
the general term of the series is of the form
$(-\e)^{n+m} R_n\big(\asb(t)/\om\big)$ and is reduced to a power series in
$\as(t)$ by repeated application of the identity
\begin{equation}\label{identity}
  -\e \as(t) \rho\Big(\frac{\asb(t)}{\om}\Big) g_{\e}^{(\MSbar)}(t)
  = \as(t) \gamma_0\Big(\frac{\asb(t)}{\om}\Big)
  \tilde{\rho}\Big(\frac{\asb(t)}{\om}\Big) g_{\e}^{(\MSbar)}(t)
  -\frac{\dif}{\dif t} \left[ \as(t)
  \tilde{\rho}\Big(\frac{\asb(t)}{\om}\Big) g_{\e}^{(\MSbar)}(t) \right] \;,
\end{equation}
valid for any function $\rho$, where
\begin{equation}\label{defD}
  \tilde{\rho} \equiv \left(1+\frac{\partial}{\partial\log\as(t)}\right)^{-1}
  \rho \equiv \left( 1 + \hat{D} \right)^{-1} \rho \;.
\end{equation}

Eq.~(\ref{identity}) means that multiplication by $-\e$ corresponds to the
operator $\gamma_0\big(\asb(t)/\om\big) (1+\hat{D})^{-1}$ after integration by
parts.  It follows that the quark anomalous dimension is given by
\begin{equation}\label{gqgMSbar}
  \gamma_{\pq\pg}^{(\MSbar)}\big(\as(t);\om\big)
  = \as(t) \qResR\bigg( -\gamma_0\Big(\frac{\asb(t)}{\om}\Big)
  \frac1{1+\hat{D}} \bigg) \sum_{n=0}^{\infty} \left(
  \gamma_0\Big(\frac{\asb(t)}{\om}\Big) \frac1{1+\hat{D}} \right)^n
  R_n\Big(\frac{\asb(t)}{\om}\Big) \;,
\end{equation}
which is the resummed expression we were looking for.

The subtlety of the result~(\ref{gqgMSbar}) is that, in order to get a resummed
formula in $\asb(t)/\om$ of the $\e$-independent $\gamma_{\pq\pg}$, we need an
all order $\e$-expansion of the $\kt$-factorisation formula~(\ref{gqgxgMSbar}).
The fact that terms of order $\e^n$ generate finite contributions to both
coefficient and anomalous dimension is somewhat similar to what already noticed
in the gluon channel, and is typical of the minimal subtraction recipe. However,
while higher orders in $\e$ correspond to higher subleading $\log1/x$
resummation levels in the gluon case, here they just correspond to higher orders
in the $\asb(t)/\om$-expansion of $\gamma_{\pq\pg}$.

In order to use Eq.~(\ref{gqgMSbar}) we need to understand the action of the
$\gamma_0 (1+\hat{D})^{-1}$ operator. A simple example is provided by setting
$R_n = \delta_{n0}$ and $\gamma_0=\asb(t)/\om$. By noting that
$\hat{D} [\as(t)]^n = n [\as(t)]^n$, we obtain
\begin{equation}\label{rational}
  \gamma_{\pq\pg}^{(\MSbar)} = \frac{\as(t)}{2\pi} \sum_{n=0}^{\infty}
  \frac{\qRes_n}{n!} \left(\frac{\asb(t)}{\om}\right)^n = \as(t)
  \qResR_B\Big(-\frac{\asb(t)}{\om}\Big) = \frac{\as(t) T_R}{2} \left(
  \esp{2\frac{\asb(t)}{\om}} + \frac1{3}\esp{\frac{2}{3}\frac{\asb(t)}{\om}}
  \right) \;,
\end{equation}
where $\qRes_B$ denotes the Borel transform of $\qResR$, which is simply obtained
from Eq.~(\ref{qKernelEps}). As noticed in~\cite{CaHa94}, Eq.~(\ref{rational})
provides all rational coefficients occurring in the resummed formula for
$\gamma_{\pq\pg}^{(\MSbar)}$.

It is also possible to provide an expression for~(\ref{gqgMSbar}) involving only
quadratures, provided the functions $R_n$ are given to all orders. In
App.~\ref{a:gqg} the result is derived in terms of the intermediate function
\begin{equation}\label{interm}
  \hat{R}\Big(\frac{\asb}{\om}\Big) = \frac{\partial}{\partial(\asb/\om)}
  \int_0^{\asb/\om} \dif a\; R_B\Big( a,
  -\int_{a}^{\asb/\om} \frac{\dif a'}{a'} \gamma_0(a') \Big) \;
\end{equation}
where $R_B$ is the Borel transform of $\qResT(\e)R(\asb/\om,\e)$ in the $-\e$
variable. The final result is
\begin{align}
  \gamma_{\pq\pg}^{(\MSbar)} = \frac{\as(t)}{2\pi}\frac{T_R}{2}
  \frac{\partial}{\partial(\asb/\om)}
  \int_0^{\asb/\om} \dif a & \left[
  \exp\left(2\int_{a}^{\asb/\om} \frac{\dif a'}{a'}\;\gamma_0(a')\right)\right.
 \nonumber \\
  &\left.+\frac1{3}\exp\left(\frac{2}{3}\int_{a}^{\asb/\om}
  \frac{\dif a'}{a'}\; \gamma_0(a')\right) \right] \hat{R}(a) \;,
 \label{gqgMSbarQuad}
\end{align}
where we notice the exponentials of~(\ref{rational}) occurring in a more general
framework.

\section{Discussion\label{s:disc}}

The main results of this paper are the renormalisation group (RG) factorisation of
the BFKL equation in $4+2\e$ dimensions at NL$x$ level, and the relation of the
$Q_0$-scheme to the $\MSbar$-scheme at the same level of accuracy. The collinear
factorisation has been proved by solving the BFKL equation with a cutoff
$Q_0^2 = \mu^2 \esp{-T}$, and by showing that --- in the on-shell limit for the
initial gluon ($Q_0=0$) --- the solution admits a RG representation with a
well-defined anomalous dimension $\bar{\gamma}_{\e}(\asb;\om)$, as given by
Eqs.~(\ref{gFormB}), (\ref{gluonEps2}) and (\ref{anDimExp}) at various levels of
accuracy. This representation shows exponentiated IR poles
$\sim\frac1{\e}\left(\frac{b\asb}{\e}\right)^n$, which are factorised in the
IR-free regime with $\asb(t) < \e/b$ and $\asb(-\infty) = 0$. Then, by the
$\e$-expansion of $\bar{\gamma}_{\e}(\asb;\om)$ in Eqs.~(\ref{gammaEps2}) and
(\ref{gammaBFKL}), we are able to define the minimal subtraction scheme, to
switch to the UV-free regime, and to find the transformation factor $R(\asb;\om)$
at NL$x$ level.

We discover in this way that the coefficient $R$ is due to the product of a
fluctuation factor $\ff$, which can be calculated at $\e=0$, and of a dynamical
factor $\rk = R/\ff$ which reflects the $\e$-dependence of
$\bar{\gamma}_{\e}(\asb;\om)$, expanded around $\e = b \asb$. Furthermore, the
$\e$-dependent $\bar{\gamma}$ directly provides subleading contributions to the
$\MSbar$-scheme anomalous dimension, encoded in
$\bar{\gamma}_{\e=b\asb}(\asb;\om)$ (Eqs.~(\ref{gammaMSbar2}), (\ref{logrk}) and
(\ref{gMSbarNL})). In this way, the difference
$\gamma^{(\MSbar)}-\gamma^{(Q_0)}$ between the two schemes is here calculated up
to NNL$x$ level. Therefore, the $\e$-dependence of the kernel is transmuted into
a subleading $\as$-dependence for the $\MSbar$ anomalous dimension.

A similar transmutation phenomenon occurs in the case of quark-gluon mixing. In
addition, due to the different form of the scheme-changing
transformation~(\ref{cff}), the $\e$-dependence induces {\em leading}
$\as/\om$-dependence into $\gamma_{\pq\pg}^{(\MSbar)}$ --- which has to be
disentangled from the process-dependent contributions to $C_{\pq\pg}^{(p)}$.
Therefore, the knowledge of the $\e$-dependence becomes increasingly important
for the full singlet evolution.

The above results are not directly applicable to the doubly resummed
approach~\cite{CCSSkernel} nor, as far as we understand, to that of~\cite{ABF}.
However, they provide some hints towards an improved scheme-changing
transformation. First of all, the analysis of Sec.~\ref{s:ugqg} and
App.~\ref{a:cfqk} shows that there is a universal part in the $\e$-dependence of
$\qcf(\gamma)$ (the $p$-scheme defining kernel) which is encoded in the
collinear pole at $\gamma = -\e$, which in turn comes from an off-shell
generalisation of the well-known $P_{\pq\pg}$ splitting function. It is then
conceivable that a similar analysis can be performed for resummed models as
well, in order to control the leading part of the scheme-changing
transformation and its mixing.

Furthermore, since the higher order $\e$-dependence affects $R$ at subleading
level, it is conceivable that most of the normalisation change comes from the
already known (or generalised) leading part. A preliminary analysis in this
direction is under way~\cite{schemes}. Hopefully, this will soon lead to a
realistic comparison of the resummed approach to experimental data.

\section*{Acknowledgements}

We are grateful to Gavin Salam and Anna Sta\'sto for a variety of discussions
and suggestions, and to Stefano Catani for quite useful discussions at various
stages of this work.

This work has been supported in part by MIUR (Italy).

\appendix
\section*{Appendices}

\section{Solutions of the difference equations in
  $\boldsymbol{\gamma}$-space\label{a:sde}}

\subsection{Asymptotic solution\label{aa:as}}

In Sec.~\ref{s:grLL} we showed that the solution of both the homogeneous and
inhomogeneous equation~(\ref{fde}) in $\gamma$-space are determined by
difference equations whose structure is, in both cases, given by (cf.\
Eqs.~(\ref{lagEq}) and (\ref{rhoEq}))
\begin{equation}\label{sfde}
  S_{\e}(\gamma+\e) - S_{\e}(\gamma) = L_{\e}(\gamma) \;.
\end{equation}
Evidently, Eq.~(\ref{sfde}) determines $S_{\e}(\gamma)$ up to an additive
periodic function of $\gamma$ of period $\e$.  Here we want to provide a
solution for $S_{\e}$ in terms of the known function $L_{\e}$. Due to the
non-local form in $\gamma$ of Eq.~(\ref{sfde}), we expect $S_{\e}$ to be
determined by an infinite number of $\gamma$-derivatives of $L_{\e}$. In fact,
$\partial_\gamma$ is the generator of $\gamma$-translations, and
$\exp(\e\partial_\gamma)$ the translation operator:
\begin{equation}\label{trans}
  \big[\exp(\e\partial_\gamma) S_{\e}\big](\gamma) = S_{\e}(\gamma+\e) \;,
\end{equation}
so that, by substituting Eq.~(\ref{trans}) into Eq.~(\ref{sfde}) we obtain the
formal solution
\begin{equation}\label{ofde}
  S_{\e} = \big[\exp(\e\partial_\gamma) - 1\big]^{-1} L_{\e} \;.
\end{equation}
Since we are not able to provide the resolvent in Eq.~(\ref{ofde}) in closed
form, we look for a perturbative solution in $\e$ of $S_{\e}$. Given that
$\exp(\e\partial_\gamma)-1\sim\e\partial_\gamma$ for $\e \to 0$, we can
rewrite the r.h.s.\ of Eq.~(\ref{ofde}) in terms of an operator close to the
identity by applying $\e\partial_\gamma$ to the left, obtaining
\begin{equation}\label{a:lagExp}
  \lag \equiv \e\partial_\gamma S_{\e} =
  \frac{\e\partial_\gamma}{\exp(\e\partial_\gamma)-1} L_{\e}
  = \sum_{n=0}^{\infty} \frac{B_n}{n!} \e^n L_{\e}^{(n)}(\gamma) \;,
\end{equation}
where we have adopted the definition~(\ref{hForm}) and used the generating
function~(\ref{bernoulli}) of the Bernoulli numbers. This proves
Eq.~(\ref{lagExp}).

A remark concerning the character of the Series in Eq.~(\ref{a:lagExp}): since
the asymptotic behaviour of the Bernoulli numbers is $B_n\sim n!$, it turns out
that the series has a non-vanishing radius of convergence if and only if the
series $\sum_n L_{\e}^{(n)}(\gamma) x^n / n!$ has infinite radius of
convergence, i.e., $L_{\e}(\gamma)$ is holomorphic in the whole $\gamma$-plane.
If $L_{\e}(\gamma)$ has at least one singularity at finite values of $\gamma$,
then the series in Eq.~(\ref{a:lagExp}) is an asymptotic series of
$\lag(\gamma)$ for $\e \to 0$.

A straightforward example is provided by the Euler $\psi$ function which
satisfies $\psi(z+1)-\psi(z)=1/z$. By rescaling $z\equiv\gamma/\e$ and
$\psi(z)\equiv S_{\e}(\gamma)$, we reproduce Eq.~(\ref{sfde}) with
$L_{\e}(\gamma)=\e/\gamma$, hence
$L_{\e}^{(n)}(\gamma)=\e(-1)^n n! /\gamma^{n=1}$ which, substituted in
Eq.~(\ref{a:lagExp}) yields
$\psi'(z)=\lag(\gamma)=\sum_{n=0}^\infty B_n (-1)^n (\e/\gamma)^{n+1}$ and
finally $\psi(z)=\log z + \sum_{n=1}^\infty B_n (-1)^{n+1}/(n z^n) + c$, which
is exactly the asymptotic expansion of the $\psi$ function for large $|z|$,
provided we set the constant of integration $c=0$. Note that also the function
$z\mapsto\psi(1-z)$ satisfies the same difference equation as $\psi(z)$, and in
fact they differ by the periodic function $\pi\cotg(\pi z)$. The asymptotic
expansion of $\psi(1-z)$ differs from that of $\psi(z)$ by the constant
$-\ui\pi\sign\big(\Im(z)\big)$ and is therefore still consistent with our series
representation. In practice, this kind of ambiguities requires the choice of a
Riemann sheet for the asymptotic solution $S_{\e}(\gamma)$, which has one or
several branch-points. This is done on the basis of the regularity requirements
for the BFKL solution in the UV or IR region. For instance, the saddle
point~(\ref{saddle2}) is dominant in the Riemann sheet relevant for the UV
regular solution of type~(\ref{gammaRep}).

\subsection{Iterative solution\label{aa:is}}

The BFKL equation can be solved through the iterative method, both in $t$- and
$\gamma$-space. In this section we want to prove the equivalence of these two
procedures.

Let us first review the iterative method in $t$-space. From the integral
equation~(\ref{BFKLeq}) for $\ugd_{\e}(\kt)$ we derive the corresponding
equation for $\widetilde{\ugd}_{\e}(t) \equiv \widetilde{\ugd}_{\e}(\kt)$:
\begin{equation}\label{inteq}
  \widetilde{\ugd}_{\e}(t) = \frac{\asbmu}{\om} \esp{\e t} \left[
  1 + \int\dif t'\; K_{\e}(t-t') \widetilde{\ugd}_{\e}(t') \right] \;,
\end{equation}
where the rescaled and azimuthally averaged kernel
\begin{equation}\label{adKernel}
  K_{\e}(t-t') \equiv \frac{\esp{\e\psi(1)}}{\Gamma(1+\e)} 
  \frac{\pi \kt^2}{g^2 N_c} \ave{\Kernel(\kt,\kt')}
  \equiv \frac{\esp{\e\psi(1)}\kt^2}{2\pi^{\e} g^2 N_c} \int_{\mathbb{S}_{1+2\e}}
  \dif^{1+2\e}\hat{\kt}'\; \Kernel(\kt,\kt')
\end{equation}
is a dimensionless scale-invariant kernel whose eigenfunctions are exponentials
$\esp{\gamma t} : 0 < \Re\gamma < 1$, and the corresponding eigenvalue function
is
\begin{equation}\label{eigenvalue}
  \chie(\gamma) = \esp{-\gamma t}\int\dif t'\; K_{\e}(t-t') \esp{\gamma t'}
  \;, \qquad(0 < \Re\gamma < 1) \;.
\end{equation}
The iterative solution of Eq.~(\ref{inteq}) is then given by
\begin{equation}\label{itsolt}
  \widetilde{\ugd}_{\e}(t) = \sum_{n=0}^\infty \widetilde{\ugd}_{\e}^{[n]}(t)\;,
\end{equation}
where
\begin{subequations}\label{fiter}
\begin{align}
  \widetilde{\ugd}_{\e}^{[0]}(t) &= \frac{\asbmu}{\om}\esp{\e t} \;,
\label{f0t} \\
  \widetilde{\ugd}_{\e}^{[n]}(t) &= \frac{\asbmu}{\om}\esp{\e t} \int\dif t'\;
  K_{\e}(t-t') \widetilde{\ugd}_{\e}^{[n-1]}(t') =
  \left(\frac{\asbmu}{\om}\esp{\e t}\right)^{n+1} \prod_{k=1}^{n} \chie(k\e)
\label{fn}
\end{align}
\end{subequations}
so as to yield Eq.~(\ref{powerSol}). Note however that the action of the kernel
$K_{\e}$ on $\widetilde{\ugd}_{\e}^{[n-1]} \sim \esp{n \e t}$ in Eq.~(\ref{fn})
is defined only if $n\e < 1$. For $n > 1/\e$ the functions
$\widetilde{\ugd}_{\e}^{[n-1]}$ do not belong anymore to the domain of $K_{\e}$
and the iterative procedure breaks up. Nevertheless, from a perturbative point
of view, Eq.~(\ref{powerSol}) is meaningful, because for any given order $n$ the
coefficient of $\asb(t)^{n+1}$ is well-defined and analytic in a non-empty open
interval $]0,1/n[$ of $\e$-values.

In $\gamma$-space, on the other hand, the iterative solution~(\ref{itSol}) has
been obtained by using an initial condition with an infrared cutoff $T$ --- see
Eqs.~(\ref{f0},\ref{F0}) in comparison with Eq.~(\ref{f0t}) --- so as to deal
with analytic (non symbolic) functions.%
\footnote{The Fourier transform of an exponential is a $\delta$ function.}
In order to prove the equivalence of Eq.~(\ref{powerSol}) and Eq.~(\ref{itSol}),
we rewrite the latter in the form
\begin{align}
  f_{\e}(\gamma) &= \sum_{n=0}^\infty f_{\e}^{[n]}(\gamma) \;,
\label{itSolGamma} \\
  f_{\e}^{[n]}(\gamma) &= \left(\frac{\asbmu}{\om}\right)^{n+1}
  \frac{\esp{[\gamma-(n+1)\e]T}}{\gamma-(n+1)\e}
  \prod_{k=1}^{n} \chie(\gamma-k\e) \;,
\label{fnGamma}
\end{align}
where we have used the explicit form~(\ref{f0}) of $f^{(0)}$. In performing the
inverse Fourier transform~(\ref{d:gammaRep}), the real part $\Re\gamma=c$ of
the integration path must lie to the right of the UV singularities of
$f_{\e}^{[n]}(\gamma)$ at $\gamma=m\e:m=1,\cdots,n+1$ and to the left of the IR
ones at $\gamma = 1+m\e:m=1,\cdots,n+1$. Therefore, also in $\gamma$-space the
iterative solution appears to be defined only for $n<1/\e$. Here, however, we
can deform the integration path and extend the validity of the
$\gamma$-representation to higher values of $n$, though this requires crossing
the real axis at larger and larger values of $\gamma>n\e$ and to properly
exclude the IR poles of the eigenvalue functions $\chie(\gamma-m\e)$ for
$\gamma-m\e>1$.

We thus calculate, for $n<1/\e$ and $(1+n)\e < c < 1+\e$
\begin{equation}\label{fnRep}
 \widetilde{\ugd}_{\e}^{[n]}(t) = \int_{\Re\gamma=c}\frac{\dif\gamma}{2\pi\ui}\;
 \esp{\gamma t} f_{\e}^{[n]}(\gamma) = \left(\frac{\asbmu}{\om}\right)^{n+1}
 \esp{-(n+1)\e T} \int_{\Re\gamma=c}\frac{\dif\gamma}{2\pi\ui} \;
 \frac{\esp{\gamma(t+T)}}{\gamma-(n+1)\e} \prod_{k=1}^{n}
 \chie(\gamma-k\e) \;.
\end{equation}
For $t>-T$ we can close the contour path to the left, and pick up the rightmost
pole of $f^{(0)}$ at $\gamma=(n+1)\e$ and other poles of the $\chie$'s at
$\gamma=m\e : m < n+1$. Each pole contributes with a $t-$ and $T$-dependent
exponential of the form $\exp[-(n+1-m)\e T + m\e t] : m \leq n+1$. Therefore,
the residue of the rightmost pole ($m=n+1$) is $T$-independent, while the other
residues are suppressed as integer powers of $\exp(-\e T)$ and vanish in the
$T \to +\infty$ limit. In conclusion
\begin{equation}\label{fnt}
  \widetilde{\ugd}_{\e}^{[n]}(t) = \left(\frac{\asb(t)}{\om}\right)^{n+1}
  \prod_{k=1}^{n}\chie\big((n+1-k)\e\big) + \ord{\esp{-\e T}}
\end{equation}
and, in the $T \to +\infty$ limit we reproduce the $t$-space
iteration~(\ref{fn}).

\subsection{Calculation of the $\boldsymbol{\rho}$-function\label{aa:rho}}

The calculation of the function $\rho_{\e}(\gamma)$ obeying Eq.~(\ref{rhoEq})
requires a special care with respect to the calculation of $S_{\e}(\gamma)$ in
Eq.~(\ref{lagEq}), even at leading $\e$ accuracy, because the r.h.s.\ of
Eq.~(\ref{lagEq}) is finite for $\e \to 0$, while that of Eq.~(\ref{rhoEq}) has
an essential singularity. By using the expression~(\ref{hForm}) for
$h_{\e}(\gamma)$ into the r.h.s.\ of Eq.~(\ref{rhoEq}), according to
Eq.~(\ref{lagExp}) we expect
\begin{equation}\label{drhoExp}
  \e\partial_\gamma\rho_{\e}(\gamma) = \sum_{n=0}^\infty \frac{B_n}{n!} \e^n
  \partial_\gamma^n \left[ f^{(0)}(\gamma)\esp{-L_{\e}(\gamma)}
  \esp{-\frac1{\e}\int^\gamma \lag(\gamma')\dif\gamma'} \right] \;.
\end{equation}
In computing the logarithmic $\gamma$-derivative of the product in square
brackets, the large contributions are represented by $T$ (from the exponent of
$f^{(0)}$, cf.\ Eq.~(\ref{f0})), and by the derivative of the ``action'', which
is of $\ord{1/\e}$. The other contributions are $T$-independent and regular in
$\e$, so that for higher derivatives one gets
\begin{equation}\label{dgammaEss}
  \partial_\gamma^n \left[ f^{(0)}(\gamma) \esp{-L_{\e}(\gamma)}
  \esp{-\frac1{\e}\int^\gamma \lag(\gamma')\dif\gamma'} \right]
  = f^{(0)}(\gamma)\esp{-L_{\e}(\gamma)}
  \esp{-\frac1{\e}\int^\gamma \lag(\gamma')\dif\gamma'}
  \left( T - \frac1{\e} \lag(\gamma) \right)^n \times \big[1+\ord{\e}\big] \;.
\end{equation}
Eq.~(\ref{drhoExp}) becomes then
\begin{align}
  \e\partial_\gamma\rho_{\e}(\gamma) &= f^{(0)}(\gamma)\esp{-L_0(\gamma)}
  \esp{-S_{\e}(\gamma)} \sum_{n=0}^\infty \frac{B_n}{n!}
  \left[\e T-L_0(\gamma)\right]^n \times \big[1+\ord{\e}\big]
\nonumber \\
 &= f^{(0)}(\gamma)\esp{-L_0(\gamma)} \esp{-S_{\e}(\gamma)}
 \frac{\e T-L_0(\gamma)}{\esp{\e T-L_0(\gamma)}-1}
 \times \big[1+\ord{\e}\big] \;,
\label{drhoEps}
\end{align}
where in the last equality use has been made of Eq.~(\ref{bernoulli}). Finally,
after integration in $\gamma$, one ends up with Eq.~(\ref{rho}).

\section{Characteristic functions of the quark kernels $\boldsymbol{\qKernel}$
  \label{a:cfqk}}

In this section we want to compute the characteristic functions of the
$\pg\to\pq$ kernels used to define the $\MSbar$ and $Q_0$ quark densities in the
high-energy regime according to Eq.~(\ref{defQuark}). These kernels are
naturally provided in $(x,\kt)$-space, and are related to their $\om$-moments in
Eqs.~(\ref{defQuark},\ref{defQuarkEigenv}) through a standard Mellin transform.
In the collinear limit, they turn out to be all determined by the same off-shell
generalisation of the $P_{\pq\pg}$ splitting function.

\subsection{The Catani-Hautmann kernel $\boldsymbol{\qKer^{(\mathrm{CH})}}$
        \label{aa:msbar}}
      
The $\pg\to\pq$ kernel used to define the quark density in the high-energy
regime is given in Eq.~(4.8,4.9,C.6) of Ref.~\cite{CaHa94} (we call $\xi$
their variable $z$):
\begin{equation}\label{Kqg}
 \as(t) \qKer_{\e}^{(\mathrm{CH})}\Big(\xi,\frac{Q^2}{\kt^2}\Big)
  = \left. \frac1{\xi} \hat{K}_{\pq\pg}\Big(\xi,\frac{\kt^2}{Q^2}\Big)
  \right|_{\text{Ref.~\cite{CaHa94}}}
\end{equation}
where $t=\log Q^2/\mu^2$ along this section.  We want to compute the double
Mellin transform (cf.\ Eq.~(\ref{defQuarkEigenv}))
\begin{equation}\label{defMelHMSbar}
  \qcf[\mathrm{CH}](\om,\gamma) \equiv \gamma(\gamma+\e)
  \int_0^1 \dif \xi \; \xi^\om
  \int\frac{\dif\kt^2}{\kt^2}\left(\frac{\kt^2}{Q^2}\right)^\gamma 
  \qKer_{\e}^{(\mathrm{CH})}\Big(\xi,\frac{Q^2}{\kt^2}\Big) \;.
\end{equation}
The result at $\om=0$ can be read from Eq.~(C.9) of Ref.~\cite{CaHa94} and in
our notations is
\begin{equation}\label{qcfMSbarOmZero}
 \qcf[\mathrm{CH}](\gamma) \equiv \qcf[\mathrm{CH}](0,\gamma) =
 \frac{T_R}{2\pi} (1+\e)(4-3\gamma+\e)
 \frac{\esp{\e\psi(1)}\Gamma(1+\e)\Gamma(1+\gamma)\Gamma(1-\gamma)}
 {\Gamma(1+\gamma+\e)\Gamma(4-\gamma+\e)} \;.
\end{equation}
The coefficient of the collinear pole at $\gamma=-\e$ is
\begin{equation}\label{He}
 \qRes(\e) \equiv \qcf[\mathrm{CH}](-\e) = \frac{T_R}{2\pi} \, \frac{2}{3} \,
 \frac{1+\e}{(1+2\e)(1+\frac{2}{3}\e)} \,
 \frac{\esp{\e\psi(1)}\Gamma^2(1+\e)\Gamma(1-\e)}{\Gamma(1+2\e)}   \;.
\end{equation}

\subsection{The transverse kernel $\boldsymbol{\qKer^{(T)}}$
        \label{aa:tk}}

\begin{figure}[htbp]
  \centering
  \includegraphics[height=0.2\textheight]{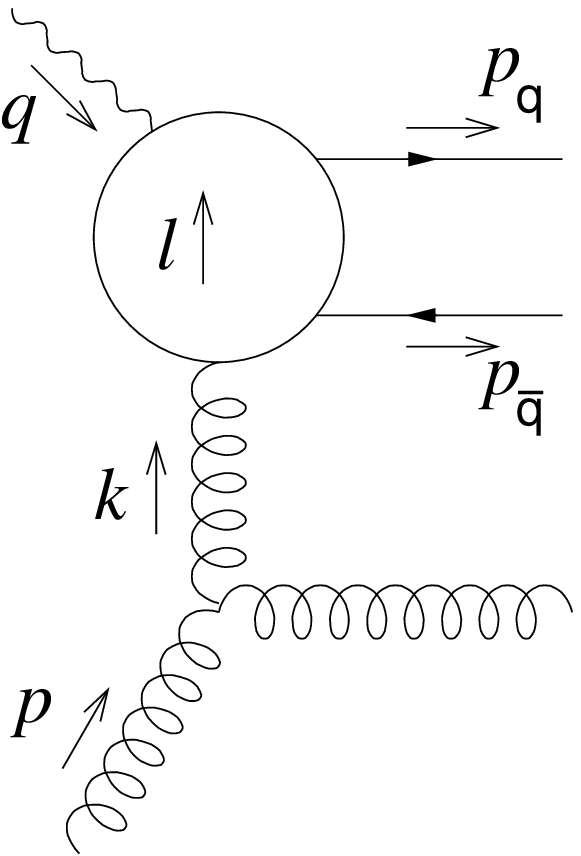}
  \caption{The process $\gamma^* \pg \to \pq \bar{\pq} \pg$ used for
         computing $\qcf[T]$.}
  \label{fig:fg_qqg}
\end{figure}

We define the $Q_0$ quark density as the high-energy limit of the quark density
in the DIS-scheme, i.e., $q^{(Q_0)} \equiv F_{2}|_{\text{high energy}}$,
the corresponding process being depicted in Fig.~\ref{fig:fg_qqg}. We choose a
reference frame in which the incoming momenta $q$ and $p$ define the
longitudinal plane. Then, in the high energy limit, the momenta entering the
blob can be decomposed as follows:
\begin{equation}\label{frame}
  q = n - xp \;; \quad k = zp + \kt \,,
\end{equation}
where $n$ and $p$ are light-like momenta spanning the longitudinal plane
orthogonal to the transverse plane. In the latter we use the euclidean
representation $\kt_\mu\kt^\mu \equiv -\kt^2 < 0$. The remaining momenta have
the following Sudakov decomposition:
\begin{subequations}\label{sudakov}
\begin{align}
  l &= p_{\pq} - q = k - p_{\bar{\pq}} = -\beta n + \alpha p + \lt \;,
 \label{l} \\
  p_{\pq} &= q + l = (1-\beta)n + (\alpha-x)p + \lt \;,
 \label{pq} \\
  p_{\bar{\pq}} &= k - l = \beta n + (z-\alpha)p + \kt - \lt \;.
 \label{pqbar}
\end{align}
\end{subequations}
We then decompose the structure function $F_2$ in transverse and longitudinal
part, each of which is given the factorisation formula~(\ref{defQuark}) with an
appropriate kernel
\begin{align}\label{Hi}
  \as(t)\qKer_{\e}^{(i)}\Big(\xi,\frac{Q^2}{\kt^2}\Big)
  &= \frac{g^2 T_R}{(2\pi)^{3+2\e}} \frac{Q^2}{\xi\kt^2} \int_0^1\dif\beta
  \int\dif^{2+2\e}\lt\;
\\ \nonumber
  &\quad \times \delta \left[1-\xi \left(1+\frac{\lt^2}{(1-\beta)Q^2}+
  \frac{(\kt-\lt)^2}{\beta Q^2} \right)\right] C_i(\beta,Q,\lt,\kt) \;,
  \qquad (i=L,T) \;,
\end{align}
where $\xi \equiv x/z$ and
\begin{align}
  C_L &= 8 \beta^2 (1-\beta)^2 Q^2 \left[\frac1{D(\lt)}-\frac1{D(\kt-\lt)}
  \right]^2
 \label{CL} \\
  C_T &= P_{\pq\gamma}(\beta) \left[\frac{\lt}{D(\lt)}-\frac{\lt-\kt}{D(\lt-\kt)}
  \right]^2
 \label{CT} \\
  D(\lt) &= \lt^2 + \beta(1-\beta)Q^2 \;; \quad
  P_{\pq\gamma}(\beta) = \frac{\beta^2 + (1-\beta)^2 +\e}{1+\e} \;.
\nonumber
\end{align}

The Mellin transform are again computed according to the
definition~(\ref{defMelHMSbar}). The first integration in $\xi$ is easily done
by exploiting the $\delta$-function:
\begin{equation}
  \as(t)\qcf[i](\om,\gamma) = 
  \frac{g^2 T_R \gamma(\gamma+\e)}{(2\pi)^{3+2\e}}\int\dif\beta\,\dif^{2+2\e}\lt\,
  \frac{\dif\kt^2}{\kt^2} \left(\frac{\kt^2}{Q^2}\right)^{\gamma-1}
  \left(1+\frac{\lt^2}{(1-\beta)Q^2}+
  \frac{(\kt-\lt)^2}{\beta Q^2} \right)^{-\om} C_i \;.
\end{equation}
The case $\e=0$ has been computed exactly in Ref.~\cite{BiNaPe01}:
\begin{equation}\label{qcfEpsZero}
 \qRes_{0}^{(i)}(\om,\gamma) =
 \frac{T_R \gamma^2}{2\pi^2\alpha_{\mathrm{em}}\as}
 \left. S_i(N=\om,\gamma)\right|_{\text{Ref.~\cite{BiNaPe01}}} \;.
\end{equation}
At $\e\neq0$ we are mainly interested in the $\om=0$ expression of the
transverse kernel, which provides the singular anomalous dimension behaviour
\begin{align}
  \as(t)\qcf[T](\gamma) &\equiv \as(t)\qcf[T](0,\gamma)
 \label{qcfTrOmZero}\\
  &= \frac{g^2 T_R \gamma(\gamma+\e)}{(2\pi)^{3+2\e}}\int_0^1\dif\beta \;
  P_{\pq\gamma}(\beta) \int\frac{\dif\kt^2}{\kt^2}
  \left(\frac{\kt^2}{Q^2}\right)^{\gamma-1}
  \int\dif^{2+2\e}\lt\;\left[\frac{\lt}{D(\lt)}-\frac{\lt-\kt}{D(\lt-\kt)}\right]^2 \;.
  \nonumber
\end{align}
The $\lt$ integral can be performed by expanding the square, shifting
integration variable $\lt-\kt\mapsto\lt$ in the square of the second term inside
the brackets, collecting all terms into a single fraction and using Feynman's
parametrisation of the common denominator
\begin{align}\label{FeynPar}
  \frac1{D^2(\lt)D(\lt-\kt)} = \int_0^1\dif\lambda\;
  \frac{2(1-\lambda)}{(\pt^2+M^2)^3} \;, \quad  \pt \equiv \lt - \lambda\kt \;,
  \quad M^2 \equiv \beta(1-\beta)Q^2 + \lambda(1-\lambda)\kt^2 \;.
\end{align}
The corresponding numerator expressed in terms of the new variables reads
\begin{equation}\label{numerator}
  (1-\lambda)\kt^2 \pt^2 - 2\lambda(\kt\cdot\pt)^2 + \lambda\kt^2 M^2 +
  \text{ odd terms in } \pt \;.
\end{equation}
Then, by means of standard integrals in $2+2\e$ dimensions, one can perform the
$\pt$ integrals obtaining additional $\beta$- $\lambda$- and $\kt^2$-dependent
factors. One can now perform the $\kt^2$ integration
\begin{equation}\label{ktIntegral}
  \int\frac{\dif\kt^2}{\kt^2}\left(\frac{\kt^2}{Q^2}\right)^{\gamma-1}
  \frac{\kt^2}{M^{2(1-\e)}} = Q^{2\e} B(\gamma,1-\gamma-\e)
  \frac{[\beta(1-\beta)]^{\gamma+\e-1}}{[\lambda(1-\lambda)]^\gamma} \;,
\end{equation}
which provides, among other things, the $\e$-running of the coupling.
Then we evaluate the $\lambda$ integral
\begin{equation}\label{lambdaIntegral}
  \int_0^1\dif\lambda\;(1-\lambda)[(1+\e)(1-\lambda)-\e\lambda]
  [\lambda(1-\lambda)]^{-\gamma} =
  \frac{\Gamma^2(1-\gamma)}{\Gamma(2-2\gamma)}\,\frac{2+\e-\gamma}{2(3-2\gamma)}
\end{equation}
and the $\beta$ integral
\begin{equation}\label{betaIntegral}
  \int_0^1\dif\beta P_{\pq\gamma}(\beta) [\beta(1-\beta)]^{\gamma+\e-1} =
  \frac{\Gamma(1+\gamma+\e)}{\Gamma(2+2\gamma+2\e)}
  \frac{2[\gamma(1+2\e)+1+2\e+2\e^2]}{(1+\e)(\gamma+\e)}
\end{equation}
obtaining the final result
\begin{align}
 \qcf[T](\gamma) = \frac{T_R}{\pi} \,&
 \frac{\esp{\e\psi(1)}\Gamma(1+\gamma)\Gamma^2(1-\gamma)\Gamma^2(1+\gamma+\e)
 \Gamma(1-\gamma-\e)} {\Gamma(2-2\gamma)\Gamma(2+2\gamma+2\e)}
\nonumber \\
  & \times\frac{(2-\gamma+\e)[\gamma(1+2\e)+1+2\e+2\e^2]}{(1+\e)(3-2\gamma)} \;.
\label{Ht}
\end{align}
The coefficient of the collinear pole of $\qcf[T](\gamma)$ at $\gamma=-\e$
equals twice the analogous quantity $\qRes(\e)$ of Eq.~(\ref{He}) obtained from
the Catani-Hautmann kernel, because here we consider both the quark and antiquark
contributions in the fermion loop, while in the kernel $\hat{K}_{\pq\pg}$ given
in~(\ref{Kqg}) only the quark (or antiquark) contributes.

\subsection{Universal collinear behaviour and off-shell splitting
  function\label{aa:ucb}}

The fact that the residue at $\gamma = -\e$ of $\qcf(\gamma)/\gamma(\gamma+\e)$
is independent of $p$ is a simple consequence of the universal collinear
behaviour of $\qKernel$ in the region $\lt^2 \sim \kt^2 \ll Q^2$, when either
$l$ or $k-l$ are collinear with $k$. In this region we can solve the $\delta$
function in Eq.~(\ref{Hi}) for small values of $\beta$, namely
\begin{equation}\label{smallBeta}
  \beta = \frac{\xi}{1-\xi} \frac{(\kt-\lt)^2}{Q^2} \ll 1 \;,
\end{equation}
or instead in the symmetrical region of $1-\beta$ small, with $\lt$ replaced by
$\kt-\lt$. Taking the region (\ref{smallBeta}), which corresponds to a collinear
quark, we see that the longitudinal part is power suppressed, so that it plays
no role in the following, and does not contribute to the $\gamma = -\e$ pole.
Furthermore, we can set $P_{\pq\gamma}(\beta) = 1$ in the transverse part, with
the denominators
\begin{subequations}\label{denominators}
\begin{align}
 (1-\xi) D(\lt) &\simeq (1-\xi) \lt^2 + \xi (\kt-\lt)^2 = \tilde{\lt}^2 +
 \xi(1-\xi) \kt^2 \equiv \tilde{D}
\\
 (1-\xi) D(\lt-\kt) &\simeq (\kt-\lt)^2 \;, \qquad
 (\tilde{\lt} \equiv \lt-\xi\kt) \;.
\end{align}
\end{subequations}
By replacing (\ref{smallBeta}) and (\ref{denominators}) in (\ref{Hi}) we obtain
\begin{equation}\label{HTcoll}
 \as(t)\qKer_{\e}^{(T)}\Big(\xi,\frac{Q^2}{\kt^2}\Big)
  = \frac{g^2 T_R}{(2\pi)^{3+2\e}\kt^2} \int\dif^{2+2\e}\tilde{\lt} \;
 \frac{\lt^2(\kt-\lt)^2 + 2 \lt\cdot(\kt-\lt)\tilde{D}
 +\tilde{D}^2}{\tilde{D}^2} \;.
\end{equation}
Introducing the boost invariant transverse momentum
$\tilde{\lt} \equiv \lt -\xi\kt$, the numerator in Eq.~(\ref{HTcoll}) can be
recast in the form
\begin{equation}\label{numeratorT}
 \tilde{\lt}^2 \kt^2 - 4\xi(1-\xi)(\kt\cdot\tilde{\lt})^2
 + 4\xi^2(1-\xi)^2(\kt^2)^2 + 4\xi(1-\xi)(1-2\xi) \kt^2(\kt\cdot\tilde{\lt})
\end{equation}
which, after azimuthal averaging in $\tilde{\lt}$ at fixed $\tilde{\lt}^2$ and
$\kt^2$, yields ($\tilde{l}^2_{\mathrm{max}} = \lambda(\xi) Q^2$)
\begin{equation}\label{HTlim}
 \qKer_{\e}^{(T)}\Big(\xi,\frac{Q^2}{\kt^2}\Big)
 \simeq \frac{\esp{\e\psi(1)}}{2\pi\Gamma(1+\e)}
 \int_0^{\lambda(\xi)Q^2}\frac{\dif\tilde{\lt}^2}{\tilde{\lt}^2} \;\left(
 \frac{\tilde{\lt}^2}{Q^2}\right)^{\e} \hat{P}_{\pq\pg}^{(0)}\Big(
 \xi,\frac{\kt^2}{\tilde{\lt}^2},\e\Big) \;, \quad
 (\lt^2, (\kt-\lt)^2 \ll Q^2) \;.
\end{equation}
This behaviour is identical to that of the Catani-Hautmann kernel
$\qKer_{\e}^{(\mathrm{CH})}$, with the transverse momentum dependent splitting
function
\begin{equation}\label{split}
 \hat{P}_{\pq\pg}^{(0)}(\xi,\kappa^2,\e) = T_R
 \left(\frac1{1+\xi(1-\xi)\kappa^2}\right)^2 \left[
 \frac{\xi^2+(1-\xi)^2+\e}{1+\e} + 4\xi^2(1-\xi)^2 \kappa^2 \right] \;,
 \quad \Big(\kappa^2 \equiv \frac{\kt^2}{\tilde{\lt}^2}\Big) \;,
\end{equation}
which thereby acquires a universal meaning.  The corresponding contribution to
$\qcf(\om,\gamma)$ becomes
\begin{align}
 \left.\frac{\qcf(\om,\gamma)}{\gamma(\gamma+\e)}\right|_{\mathrm{coll}}
 &= \int_0^1 \dif\xi \; \xi^{\om}\int_0^\infty\frac{\dif\kt^2}{\kt^2} \;
 \left(\frac{\kt^2}{Q^2}\right)^\gamma
 \qKer_{\e}^{(T)}\Big(\xi,\frac{Q^2}{\kt^2}\Big)
\\
 &= \frac{\esp{\e\psi(1)}}{2\pi\Gamma(1+\e)} \int_0^1 \dif\xi\;
 \xi^{\om} \int_0^{\lambda(\xi)Q^2} \frac{\dif\tilde{\lt}^2}{\tilde{\lt}^2} \;
 \left(\frac{\tilde{\lt}^2}{Q^2}\right)^{\gamma+\e} \cdot \int_0^\infty
 \frac{\dif\kappa^2}{\kappa^2}\; (\kappa^2)^\gamma
 \hat{P}_{\pq\pg}^{(0)}(\xi,\kappa^2,\e)
\\
 &\stackrel{\gamma\simeq-\e}{\simeq}
 \frac1{\gamma+\e} \,\frac{\esp{\e\psi(1)}}{2\pi\Gamma(1+\e)} \int_0^1 \dif\xi\;
 \xi^{\om} \left[ \int_0^\infty \frac{\dif\kappa^2}{\kappa^2}\;
 (\kappa^2)^{\gamma} \hat{P}_{\pq\pg}^{(0)}(\xi,\kappa^2,\e)
 \right]_{\gamma=-\e}\;.
\end{align}
We see that the residue at the $\gamma = -\e$ pole is independent of the details
of the $\tilde{\lt}$ phase space $\tilde{\lt}^2 < \lambda(\xi)Q^2$ and is given by a
simple (analytically continued) moment of the $\hat{P}_{\pq\pg}^{(0)}$ splitting
function. We obtain, for $\om = 0$, the explicit expression
\begin{align}
 \qRes(\e) &= \frac{\esp{\e\psi(1)}}{2\pi\Gamma(1+\e)}
 \int_0^1\dif\xi \int_0^\infty\dif\kappa^2 \;
 (\kappa^2)^{-\e} \left[ -\frac{\partial}{\partial\kappa^2}
 \hat{P}_{\pq\pg}^{(0)}(\xi,\kappa^2,\e) \right] \;.
\\
 &= \frac{T_R}{2\pi} \, \frac23 \, \frac{1+\e}{(1+2\e)(1+\frac23 \e)} \,
 \frac{\esp{\e\psi(1)}\Gamma^2(1+\e)\Gamma(1-\e)}{\Gamma(1+2\e)}
\end{align}
The above analysis shows that, by combining $\kt$-factorisation with the
collinear behaviour it is possible to define an off-shell splitting function
which is process-independent. The residue at the pole $\gamma = -\e$ is a
particular moment of such splitting function.


\section{$\boldsymbol{\ord{\e}}$-corrections to the gluon density
  \label{a:cgd}}

In order to compute the corrections of relative order $b\asb$ to $\rk$ and to
$\gamma^{(\MSbar)}$ we have to calculate the corrections of order $\ord{\e}$
--- i.e., of relative order $\ord{\e^2}$ --- to the exponent of the solution
(\ref{gammaRepB}) of the (unintegrated) gluon density $\widetilde{\ugd}_{\e}$.

\subsection{Calculation of the saddle point fluctuations\label{aa:cspf}}

These corrections can be computed by means of a saddle point expansion of the
$\gamma$-representation
\begin{equation}\label{gammaRep2}
 \widetilde{\ugd}_{\e}(t) = M_{\e} \int\dif\gamma\;
 \esp{\gamma t + S_{\e}(\gamma)} \;.
\end{equation}
Here the ``action'' $S_{\e}$ has to be expanded to $\ord{\e}$ as done in
Eq.~(\ref{action}), and the normalisation
\begin{equation}\label{d:Me}
  M_{\e} = \sqrt{\frac{\asbmu}{2\pi\e\om}}
\end{equation}
can be inferred from Eq.~(\ref{gammaRep}), which also sets to 0 the lower bound
of the integral in the action. We split the exponent of Eq.~(\ref{gammaRep2})
into an $\e$-regular part
\begin{equation}\label{regPart}
 \tilde{E}_{\e}(\gamma) \equiv -\frac1{2} L_{\e}(\gamma)
 + \frac{\e}{12} L_{\e}'(\gamma) + \ord{\e^2}
\end{equation}
plus a leading-$\e$ phase
\begin{equation}\label{leadPhase}
 E_{\e}(\gamma) \equiv \gamma t
 + \frac1{\e}\int_0^\gamma \dif\gamma'\; L_{\e}(\gamma') \;,
\end{equation}
whose stationarity condition determines the saddle point
$\gamma=\bar{\gamma}_{\e}$ (cf.\ Eq.~(\ref{saddleEps2})), and the size of the
fluctuations
\begin{equation}\label{fluctEps}
 \sigma_{\gamma}^2 \equiv -\frac1{E''(\bar{\gamma}_{\e})}
  = \frac{\e}{-L_{\e}'(\bar{\gamma}_{\e})} \;.
\end{equation}
By expanding around the saddle point and factoring out of the integral the value
of the integrand at the saddle point, we can write
\begin{equation}\label{spExp}
  \widetilde{\ugd}_{\e}(t) = M_{\e} \esp{\bar{\gamma}_{\e} t
  + S_{\e}(\bar{\gamma}_{\e})} \sqrt{2\pi}\sigma_{\gamma}
  \ave{
    \exp\left[ \sum_{k=3}^\infty \frac{E^{(k)}(\bar{\gamma}_{\e})}{k!}
      (\gamma-\bar{\gamma}_{\e})^k
      + \sum_{k=1}^\infty \frac{\tilde{E}^{(k)}(\bar{\gamma}_{\e})}{k!}
      (\gamma-\bar{\gamma}_{\e})^k \right]
  } \;,
\end{equation}
where the ``average'' $\ave{\cdots}$ is defined by
\begin{equation}\label{d:average}
  \ave{ f(\gamma) } \equiv
  \int \frac{\dif\gamma}{\sqrt{2\pi}\sigma_{\gamma}} \;
  \exp\left[-\frac{(\gamma-\bar{\gamma}_{\e})^2}{2\sigma_{\gamma}^2}\right]
  f(\gamma) \;.
\end{equation}
In Eq.~(\ref{spExp}) the sum of the $E^{(k)}$ starts from third order because the
zeroth order has been taken out of the integral, the first order is zero by
definition of the saddle point condition, and the second order provides the
gaussian measure in the ``average integral''~(\ref{d:average}).

The product out of the ``average'' in Eq.~(\ref{spExp}) ---
by using the expressions~(\ref{d:Me},\ref{fluctEps}) for $M_{\e}$ and
$\sigma_{\gamma}$ and Eqs.~(\ref{action},\ref{d:Lb},\ref{expintgamma}) for
the exponential --- yields
\begin{equation}\label{prefactor}
 M_{\e} \, \esp{\bar{\gamma}_{\e} t + S_{\e}(\bar{\gamma}_{\e})}
 \sqrt{2\pi}\sigma_{\gamma} = \frac1{\sqrt{-\chi_{\e}(\bar{\gamma}_{\e})}}
 \exp\left[\int_{-\infty}^{t} \dif\tau\; \bar{\gamma}_{\e}(\tau) \right]
 \esp{\e\frac1{12}L_{\e}'(\bar{\gamma}_{\e})} \;,
\end{equation}
i.e., the gluon density of Eq.~(\ref{ugdTildeB}) times a correction factor
\begin{equation}\label{corr1}
  \esp{\e\frac1{12}L_{\e}'(\bar{\gamma}_{\e})}
  = 1 + \e\frac1{12}L_{\e}'(\bar{\gamma}_{\e}) + \ord{\e^2} \;.
\end{equation}
The remaining corrections are found from the ``average integral''. Note first
that the ``average'' of powers of $\gamma-\bar{\gamma}_{\e}$ vanishes for odd
powers, and is given by
\begin{equation}\label{powAverage}
 \ave{ (\gamma-\bar{\gamma}_{\e})^{2n} }
 = (2n-1)!! \, \sigma_{\gamma}^{2n}
 = \e^n \frac{(2n-1)!!}{\big[-L_{\e}'(\bar{\gamma}_{\e})\big]^n}
\end{equation}
for even powers, thus providing higher and higher orders in $\e$ as $n$
increases. However, the coefficients of the saddle point expansion present
$\e$-singularities, due to the leading phase derivatives $E^{(k)} \sim 1/\e$.
When one expands the exponential inside the ``average'', the $m$-th order (which
is of $\ord{1/\e^m}$) contains powers $(\gamma-\bar{\gamma}_{\e})^j$ only for
$j \geq 3m$. Therefore the power in $\e$ of the fluctuations~(\ref{powAverage})
increases faster than the inverse power in $\e$ of the order of expansion of the
exponential. In practice, if one is interested in computing the ``average'' to,
say, $l$-th order in $\e$, it suffices to expand the exponential up to order
$m \leq 2 l$ and to take terms in the sum up to $k \leq 2 l + 3 - m$.
%
In the present case, we are interested to the first $\e$-corrections, namely
$l=1$, and we need therefore the first order expansion of the exponential up to
$k=4$ and the second order up to $k=3$. By denoting
$\delta \equiv (\gamma-\bar{\gamma}_{\e})$ we have
\begin{align}
  \ave{\cdots}_{\text{\ref{spExp}}} &= \ave{
    1 + \frac{\tilde{E}^{(2)}}{2!} \delta^2
    + \frac{E^{(4)}}{4!} \delta^4
    + \frac1{2}\left(\frac{\tilde{E}^{(1)}}{1!}\delta\right)^2
    + \frac{E^{(1)}}{1!} \frac{\tilde{E}^{(3)}}{3!}
    \delta^4
    + \frac1{2}\left(\frac{E^{(3)}}{3!}\delta^3\right)^2
  } + \ord{\e^2}
\nonumber \\
  &= 1 + \e\left[\frac1{8}(-L') + \frac{5}{24}\frac{L''{}^2}{(-L')^3}
  +\frac1{8}\frac{L'''}{(-L')^2} \right] + \ord{\e^2} \;,
\label{aveExp}
\end{align}
where we have used
\begin{equation}\label{E}
  E^{(k)} = \frac{L^{(k-1)}}{\e} \;, \quad E^{(k)} = -\frac1{2} L^{(k)} +
  \ord{\e} \;,
\end{equation}
and the dependence on $\e$ and $\bar{\gamma}_{\e}$ has been omitted. Finally,
the full $\ord{\e}$ correction to the action~(\ref{S1}) is obtained
by multiplying the saddle point fluctuations of Eq.~(\ref{aveExp}) by the first
order correction to the action at the saddle point in Eq.~(\ref{corr1}).

\subsection{Cancellation of the singularities\label{aa:cs}}

The terms of the $\ord{\e}$-correction to the gluon density $S_1$ present
various singularities deriving from those of the function $L_{\e,b}'(\gamma)$
written in Eq.~(\ref{Lprime}): {\it (1)} poles at $\bar{\gamma}_{\e} = 0$, i.e.,
at $\asb(t) = 0$; {\it (2)} poles at $\chie(\bar{\gamma}_{\e}) = b\om/\e$, i.e.,
at $\asb(t) = \e/b$; {\it (3)} poles at $\e = 0$ because of the
$b\asb/\e$-expansion in the denominator of $L_{\e,b}^{(n)}$.

In order to better understand the role of these singularities, we rewrite
$S_1$ in terms of the inverse of $L_{\e,b}'$
\begin{equation}
  \label{d:invL}
  \lambda(\gamma) \equiv \frac1{L_{\e,b}'(\gamma)} =
  \frac{\chie(\gamma)-\frac{b\om}{\e}}{-\chie'(\gamma)}
\end{equation}
and of the eigenvalue functions $\chie$ and its derivatives.
By computing the derivatives of $\lambda$
\begin{equation}\label{lambdaDer}
 \lambda' = -\frac{L''}{(L')^2} = 1-\frac{\chie''}{\chie'} \lambda \;, \quad
 \lambda'' = 2 \frac{(L'')^2}{(L')^3} - \frac{L'''}{(L')^2}
 =  - \frac{\chie''}{\chie'} + \left[\left(\frac{\chie''}{\chie'}\right)^2 -
 \left(\frac{\chie''}{\chie'}\right)' \right] \lambda
\end{equation}
we deduce
\begin{equation}\label{Lratios}
 \frac{(L'')^2}{(L')^3} = \frac1{\lambda} - 2\frac{\chie''}{\chie'}
 + \lambda \left(\frac{\chie''}{\chie'}\right)^2 \;, \quad
 \frac{L'''}{(L')^2} = \frac{2}{\lambda} - 3 \frac{\chie''}{\chie'} + \lambda
 \left[\left(\frac{\chie''}{\chie'}\right)'
 + \left(\frac{\chie''}{\chie'}\right)^2\right] \;,
\end{equation}
so as to obtain
\begin{equation}\label{S1lambda}
  S_1 =  \frac1{24}\frac{\chie''}{\chie'} + \left[
 \frac1{8} \left( \frac{\chie''}{\chie'} \right)'
 -\frac1{12} \left(\frac{\chie''}{\chie'} \right)^2 \right]\lambda \;.
\end{equation}
which agrees with Eq.~(\ref{S1form}) after replacing the
expression~(\ref{d:invL}) of $\lambda$ and evaluating at
$\gamma=\bar{\gamma}_{\e}$.

\section{Expression of $\boldsymbol{\gamma_{\pq\pg}}$ by quadratures
\label{a:gqg}}

In this appendix we want to show how it is possible to give an expression for
the NL$x$ resummation of $\gamma_{\pq\pg}^{(\MSbar)}$ in terms of the universal
functions $R(\asb/\om,\e)$ and $\qRes(\e)$ up to quadratures.

We start from Eq.~(\ref{gqgMSbar}) and notice that, besides the overall factor
$\as(t)$, all the remaining dependence on $\as(t)$ occurs through the ratio
$a\equiv a(t) \equiv \asb(t)/\om$. In fact, also the operator $\hat{D}$ defined
in Eq.~(\ref{defD}) can be written as $\hat{D} = a \partial_a$.  Let us then
introduce the operator
\begin{equation}\label{defO}
 \hat{O} \equiv \gamma_0 [1+\hat{D}]^{-1} = \gamma_0(a) [ \partial_a \circ a ]^{-1} 
\end{equation}
and, by using the expansions~(\ref{Rexpns},\ref{qKexpns}), rewrite
Eq.~(\ref{gqgMSbar}) in the form
\begin{equation}\label{gqg2sums}
 \gamma_{\pq\pg}^{(\MSbar)} = \as \sum_{m=0}^\infty \qRes_m \hat{O}^m
 \sum_{n=0}^\infty \hat{O}^n R_n(a) \;.
\end{equation}

The first step is to derive an expression for arbitrary powers of $\hat{O}$. We
solve this problem by considering the resolvent of $\hat{O}$
\begin{equation}\label{resO}
 Z_{\hat{O}}(\lambda) \equiv [1-\lambda \hat{O}]^{-1}
 = \sum_{k=0}^\infty \lambda^k \hat{O}^k \;,
\end{equation}
where $\lambda$ is a complex parameter. The resolvent is the generating function
of the powers of $\hat{O}$, in the sense that
\begin{equation}\label{powO}
 \hat{O}^n = \frac1{n!}\,\left. \frac{\dif^n}{\dif \lambda^n}
  Z_{\hat{O}}(\lambda) \right|_{\lambda=0} \;.
\end{equation}
An expression for $Z_{\hat{O}}(\lambda)$
is obtained by solving the equation
\begin{equation}\label{resEq}
 y(a) = [ Z_{\hat{O}}(\lambda) x ](a) \quad \iff \quad
 x = [1-\lambda \hat{O}] y = y - \lambda \gamma_0 [1+\hat{D}]^{-1} y
\end{equation}
where $y(a)$ is the unknown function to be determined in terms of $x(a)$. By
introducing the function $w(a)$ such that $y = [1+\hat{D}]w = \partial_a(a w)$,
we obtain a first order differential equation for the function $a w(a)$ which
can be easily solved. The general solution of Eq.~(\ref{resEq}) is
\begin{equation}\label{genSol}
 [ Z_{\hat{O}}(\lambda) x ](a) =
 \partial_a \int^a \dif a' \; \exp\left[ \lambda \int_{a'}^a 
 \frac{\dif a''}{a''} \; \gamma_0(a'') \right] x(a') \;.
\end{equation}
The lower bound of the $a'$-integral is 0. It can be determined by comparing the
expression of $\hat{O}$ we obtain from Eqs.~(\ref{powO}) and (\ref{genSol})
\begin{equation}\label{genO}
 [ \hat{O} x ](a) = \left. \frac{\dif}{\dif \lambda}
 [ Z_{\hat{O}}(\lambda) x ](a) \right|_{\lambda=0}
 = \partial_a \int^a \dif a'\; x(a') \int_{a'}^a \frac{\dif a''}{a''} \;
 \gamma_0(a'') = \frac{\gamma_0(a)}{a}\int^a \dif a' \; x(a')
\end{equation}
with the integral representation of the operator $\hat{O}$ that can be obtained
by the series expansion of $x(a) = \sum_{n=0}^\infty x_n a^n$:
\begin{equation}\label{intRepO}
 [ \hat{O} x ](a) = \left[\gamma_0\frac1{1+\hat{D}} x \right](a)
 = \gamma_0(a) \sum_{n=0}^\infty \frac{x_n a^n}{1+n}
 = \frac{\gamma_0(a)}{a} \int_0^a \dif a' \; x(a') \;.
\end{equation}

The second step is to perform the sum in $n$ in Eq.~(\ref{gqg2sums}). By using
again Eqs.~(\ref{powO}) and (\ref{genSol}) we obtain
\begin{align}
 \sum_{n=0}^\infty \hat{O}^n R_n(a) &= \partial_a \int_0^a \dif a' \;
 \left[\int_{a'}^a \frac{\dif a''}{a''} \gamma_0(a'') \right]^n \frac{R_n(a')}{n!}
\nonumber \\
 &= \partial_a \int_0^a \dif a' \; R_B \Big(a',
 -\int_{a'}^a \frac{\dif a''}{a''} \gamma_0(a'') \Big) \equiv \hat{R}(a) \;,
\label{sumN}
\end{align}
where we have noticed that the sum inside the integral is just the Borel
transform of the function $\qResT(\e)R(a,\e)$ in the $(-\e)$-variable.

The third step is to perform the sum in $m$ in Eq.~(\ref{gqg2sums}). Here we
exploit the particular structure of the coefficients $\qRes_m$ of being a linear
combination of powers, since $\qResR(\e)$ (cf.\ Eq.~(\ref{qKernelEps})) is a
linear combination of simple fractions:
\begin{equation}\label{qResM}
 \qRes_m = \frac{T_R}{4\pi} \left[ 2^m + \frac13 \left(\frac23\right)^m \right] \;,
\end{equation}
so that, by using the second equality of Eq.~(\ref{resO}), we obtain just a
linear combination of resolvents with parameters $\lambda = 2$ and $\lambda = 2/3$:
\begin{equation}\label{sumM}
 \sum_{m=0}^\infty \qRes_m \hat{O}^m = \frac{T_R}{4\pi}
 \sum_{m=0}^\infty \left[ 2^m \hat{O}^m + \frac13 \left(\frac23\right)^m
 \hat{O}^m \right]
 = \frac{T_R}{4\pi} \left[ Z_{\hat{O}}(2)
   + \frac13 Z_{\hat{O}}\Big(\frac23\Big) \right]
\end{equation}

Finally, by inserting Eqs.~(\ref{sumN}) and (\ref{sumM}) in Eq.~(\ref{gqg2sums})
yields
\begin{equation}\label{gqgResummed}
 \gamma_{\pq\pg}^{(\MSbar)} = \as \frac{T_R}{4\pi} \partial_a \int_0^a\dif a'\;
 \left[ \exp\left(2\int_{a'}^{a} \frac{\dif a''}{a''}\;\gamma_0(a'')\right)
  +\frac13\exp\left(\frac{2}{3}\int_{a'}^{a} \frac{\dif a''}{a''}\;
  \gamma_0(a'')\right) \right] \hat{R}(a') \;.
\end{equation}



\begin{thebibliography}{99}


\bibitem{BFKL}
L.N.~Lipatov, Sov.\ J.\ Nucl.\ Phys. {\bf 23} (1976) 338;\\
E.A.~Kuraev, L.N.~Lipatov and  V.S.~Fadin, Sov.\ Phys.\ JETP {\bf 45} (1977) 199;\\
I.I.~Balitsky and  L.N.~Lipatov, Sov.\ J.\ Nucl.\ Phys. {\bf 28} (1978) 822;\\
L.N.~Lipatov, Sov.\ Phys.\ JETP {\bf 63} (1986) 904.

\bibitem{DGLAP}
V.N.~Gribov and L.N.~Lipatov, Sov.\ J.\ Nucl.\ Phys. {\bf 15} (1972) 438;\\
G.~Altarelli and G.~Parisi, Nucl.\ Phys.\ B {\bf 126} (1977) 298;\\
Yu.L.~Dokshitzer, Sov.\ Phys.\ JETP {\bf 46} (1977) 641.

\bibitem{Kfact}
S.~Catani, M.~Ciafaloni and F.~Hautmann,
Phys.\ Lett.\ B {\bf 242} (1990) 97;
Nucl.\ Phys.\ B {\bf 366} (1991) 135.

\bibitem{CoEl91}
J.~C.~Collins and R.~K.~Ellis,
Nucl.\ Phys.\ B {\bf 360} (1991) 3.

\bibitem{CaCiHa93}
S.~Catani, M.~Ciafaloni and F.~Hautmann,
Phys.\ Lett.\ B {\bf 307} (1993) 147.
\bibitem{CaHa94}
S.~Catani and F.~Hautmann,
Nucl.\ Phys.\ B {\bf 427} (1994) 475

\bibitem{RGvert}
V.~S.~Fadin and L.~N.~Lipatov,
JETP Lett.\  {\bf 49} (1989) 352
[Yad.\ Fiz.\  {\bf 50} (1989\ SJNCA,50,712.1989) 1141];
Nucl.\ Phys.\ B {\bf 406} (1993) 259;
Nucl.\ Phys.\ B {\bf 477} (1996) 767.\\
V.~S.~Fadin, R.~Fiore and A.~Quartarolo,
Phys.\ Rev.\ D {\bf 50} (1994) 2265;
Phys.\ Rev.\ D {\bf 50} (1994) 5893.\\
V.~S.~Fadin, R.~Fiore and M.~I.~Kotsky,
Phys.\ Lett.\ B {\bf 359} (1995) 181;
Phys.\ Lett.\ B {\bf 387} (1996) 593;
Phys.\ Lett.\ B {\bf 389} (1996) 737.\\
V.~S.~Fadin, M.~I.~Kotsky and L.~N.~Lipatov,
BUDKER-INP-1996-92, hep-ph/9704267.\\
V.~Del Duca,
Phys.\ Rev.\ D {\bf 54} (1996) 989;
Phys.\ Rev.\ D {\bf 54} (1996) 4474.

\bibitem{QQvertCC}
S.~Catani, M.~Ciafaloni and F.~Hautmann,
Phys.\ Lett.\ B {\bf 242} (1990) 97;
Nucl.\ Phys.\ B {\bf 366} (1991) 135.\\
G.~Camici and M.~Ciafaloni, Phys.\ Lett.\ B {\bf 386} (1996) 341;
Nucl.\ Phys.\ B {\bf 496} (1997) 305
[Erratum-ibid.\ B {\bf 607} (2001) 431].

\bibitem{QQvertFFFK}
V.S.~Fadin, R.~Fiore, A.~Flachi and M.I.~Kotsky,
Phys.\ Lett.\ B {\bf 422} (1998) 287.

\bibitem{CaCi97}
G.~Camici and M.~Ciafaloni,
Nucl.\ Phys.\ B {\bf 496} (1997) 305
[Erratum-ibid.\ B {\bf 607} (2001) 431]

\bibitem{FaLi98}
V.~S.~Fadin and L.~N.~Lipatov,
Phys.\ Lett.\ B {\bf 429} (1998) 127

\bibitem{CaCi98}
M.~Ciafaloni and G.~Camici,
Phys.\ Lett.\ B {\bf 412} (1997) 396
[Erratum-ibid.\ B {\bf 417} (1998) 390]
Phys.\ Lett.\ B {\bf 430} (1998) 349

\bibitem{CCS} M.~Ciafaloni, D.~Colferai and G.P.~Salam,
Phys.\ Rev.\ D {\bf 60} (1999) 114036.

\bibitem{CCSSkernel}
M.~Ciafaloni, D.~Colferai, G.P.~Salam and A.M.~Sta\'sto,
Phys.\ Lett.\ B {\bf 576} (2003) 143;
Phys.\ Rev.\ D {\bf 68} (2003) 114003.

\bibitem{ABF}
G.~Altarelli, R.D.~Ball and S.~Forte, Nucl.\ Phys.\ B {\bf 575} (2000) 313;
Nucl.\ Phys.\ B {\bf 599} (2001) 383;
Nucl.\ Phys.\ B {\bf 621} (2002) 359;
Nucl.\ Phys.\ B {\bf 674} (2003) 459;
hep-ph/0310016.

\bibitem{Q0}
M.~Ciafaloni, Phys.\ Lett.\ B {\bf 356} (1995) 74;

\bibitem{Rterms}
J.~R.~Forshaw, R.~G.~Roberts and R.~S.~Thorne,
Phys.\ Lett.\ B {\bf 356}, 79 (1995)
R.~D.~Ball and S.~Forte,
Phys.\ Lett.\ B {\bf 358}, 365 (1995)

\bibitem{omExp}
M.~Ciafaloni and D.~Colferai,
Phys.\ Lett.\ B {\bf 452} (1999) 372
M.~Ciafaloni, D.~Colferai and G.~P.~Salam,
Phys.\ Rev.\ D {\bf 60} (1999) 114036

\bibitem{BiNaPe01}
A.~Bialas, H.~Navelet and R.~Peschanski,
Nucl.\ Phys.\ B {\bf 603} (2001) 218

\bibitem{schemes}
M.~Ciafaloni, D.~Colferai, G.P.~Salam and A.M.~Sta\'sto,
to appear.

\end{thebibliography}
\end{document}